\documentclass[iop,twocolappendix]{emulateapj}
\usepackage{graphicx}
\usepackage{color}
\usepackage{amsmath,amssymb}
\usepackage{bm}

\newcommand{\gsim}{\hspace{0.3em}\raisebox{0.4ex}{$>$}\hspace{-0.75em}\raisebox{-.7ex}{$\sim$}\hspace{0.3em}}
\newcommand{\lsim}{\hspace{0.3em}\raisebox{0.4ex}{$<$}\hspace{-0.75em}\raisebox{-.7ex}{$\sim$}\hspace{0.3em}}

\shorttitle{Resolved stellar streams around NGC~4631}
\shortauthors{Tanaka et al.}

\begin{document}

\title{Resolved stellar streams around NGC~4631 from a Subaru/Hyper Suprime-Cam survey}
 
\author{Mikito Tanaka\altaffilmark{1} and Masashi Chiba}
\affil{Astronomical Institute, Tohoku University, Aoba-ku,
 Sendai 980-8578, Japan}

\and

\author{Yutaka Komiyama\altaffilmark{2}}
\affil{National Astronomical Observatory of Japan, 2-21-1
 Osawa, Mitaka, Tokyo 181-8588, Japan\\ \& \\
 The Graduate University for Advanced Studies,
2-21-1 Osawa, Mitaka, Tokyo 181-8588, Japan}

\altaffiltext{1}{mikito@astr.tohoku.ac.jp}

\begin{abstract}

We present the first results of the Subaru/Hyper Suprime-Cam (HSC) survey of the interacting galaxy system, NGC~4631 and NGC~4656. From the maps of resolved stellar populations, we identify 11 dwarf galaxies (including already-known dwarfs) in the outer region of NGC~4631 and the two tidal stellar streams around NGC~4631, named Stream SE and Stream NW, respectively. This paper describes the fundamental properties of these tidal streams. Based on the tip of red giant branch method and the Bayesian statistics, we find that Stream~SE (7.10~Mpc in Expected a posteriori, EAP, with the 90\% credible intervals of [6.22, 7.29]~Mpc) and Stream~NW (7.91~Mpc in EAP with the 90\% credible intervals of [6.44, 7.97]~Mpc) are located in front of and behind NGC~4631, respectively. We also calculate the metallicity distribution of stellar streams by comparing the member stars with theoretical isochrones on the color-magnitude diagram. We find that both streams have the same stellar population based on the Bayesian model selection method, suggesting that they originated from a tidal interaction between NGC~4631 and a single dwarf satellite. The expected progenitor has a positively skewed metallicity distribution function with $\rm [M/H]_{\rm EAP}=-0.92$ with the 90\% credible intervals of [$-1.46, -0.51$]. The stellar mass of the progenitor is estimated as $3.7 \times 10^8 M_\sun$ with the 90\% credible intervals of [$5.8 \times 10^6, 8.6 \times 10^9$]~$M_\sun$ based on the mass-metallicity relation for Local group dwarf galaxies. This is in good agreement with an initial stellar mass of the progenitor presumed in the previous $N$-body simulation. 

\end{abstract}


\section{Introduction}

The currently favored cosmology based on $\Lambda$-dominated cold dark matter ($\Lambda$CDM) predicts the hierarchical formation of large galaxies like the Milky Way and M31. Indeed, several $N$-body simulations based on this cosmology \citep[e.g.,][]{BJ05,Cooper10} succeeded to reproduce diverse tidal stellar streams with various surface brightnesses, metallicities, velocities and morphologies depending on mass accretion history of galaxies \citep[e.g.,][]{Johnston08}. Therefore, stellar streams as observed in nearby galaxies provide an important opportunity to assess the outputs of the $\Lambda$CDM model. 

The Milky Way and M31 play a special role in these studies because they offer us very detailed information on stellar streams through their resolved stars. For instance, prominent stellar streams, the Sagittarius stream in the Milky Way \citep[e.g.,][]{Ibata94,Majewski03} and Giant Southern Stream in M31's halo \citep[e.g.,][]{Ibata01,Ibata07,Tanaka10,Ibata14} are the most spectacular examples, thereby enabling us to derive their dynamical histories in light of the $\Lambda$CDM model. Recently, large ground-based telescopes such as the 8.2m Subaru telescope made it possible to resolve stellar halos of other nearby galaxies beyond the Local Group, which reveal the diversity of stellar structures of galaxy outskirts with various morphologies \citep[e.g.,][]{Mouhcine10,Tanaka11,Greggio14,Okamoto15,Crnojevic16}. Furthermore, the GHOSTS survey with the Hubble Space Telescope (HST), which has observed various field along the minor and major axes of 16 nearby disk galaxies \citep{Monachesi13}, also reveals the diversity in halo color profiles and stellar halo masses, supporting the galaxy-to-galaxy scatter in halo stellar properties, a consequence of the stochasticity inherent in the assembling history of galaxies \citep{Monachesi16,Harmsen16}. However, the pencil beam surveys with the HST such as GHOSTS have a drawback that they are less able to identify new streams and overdensities because of their narrow fields.

On the other hand, \citet{Martinez10} exploited small reflecting telescopes of no more than about 50~cm in order to study stellar halos and tidal features in more distant galaxies through their integrated light. This is because integrated surface brightness is independent of distance. Subsequently, other groups inspired by \citet{Martinez10} have adopted this observational methodology to perform a direct test of the $\Lambda$CDM cosmology \citep[e.g.,][]{vanDokkum14,Javanmardi16,Merritt16}. These authors also found the varieties in the stellar halos in more distant nearby spiral galaxies beyond the Local Volume, implying stochasticity in the accretion histories of galaxies. These integrated light surveys are complementary to the GHOSTS and surveys with large ground-based telescopes based on resolved stellar populations of galaxies.

In this work, we select NGC~4631 as another test-bed aiming for a better understanding of the varieties in stellar halos. NGC~4631 is an edge-on spiral galaxy (Sc) at a distance of about $7$~Mpc \citep[e.g.,][]{Tikhonov06,Radburn-Smith11}. It appears that its companion galaxy, NGC~4656, is interacting with NGC~4631. In fact, the outskirt of NGC~4631 is interesting as an interacting system, because the presence of its extraplanar components is reported by a variety of observations, from X-rays to radio wavebands. In particular, \citet{Rand94} shows five H$_{\rm I}$ spurs surrounding NGC~4631. \citet{SH12} discovered a UV-bright, tidal dwarf galaxy candidate in the NGC~4631/4656 galaxy group, which may be originated from extremely disturbed H$_{\rm I}$, suggesting that these objects are young with age of 200-300~Myr. However, the relation between these peculiar gas structures and resolved stellar populations in this galaxy system remains unclear. Thus, the outskirts of this galaxy probed by its resolved stars will provide a unique laboratory to study a nearby system which is violently perturbed by neighbors. In particular, we expect that such an interacting galaxy is surrounded by some stellar streams or substructures with/without H$_{\rm I}$ gas. In fact, \citet{Seth05b} found somewhat thickened disk structures by HST/ACS observations, and \citet{Martinez15} discovered stellar streams at the north-west side (Stream~NW) and the south-east side (Stream~SE) of NGC~4631 based on the integrated surface light analysis. However, they found that these streams do not overlap with any H$_{\rm I}$ components.

In this paper, we report on the resolved stellar populations of the two tidally-spreading streams around NGC~4631 using the Subaru telescope. The layout of this paper is as follows. In Section~\ref{sec:data}, we show our observations and detailed procedures of reduction and photometry for our data. In Section~\ref{sec:results}, we present the discovery of new dwarf galaxies around NGC~4631, the color-magnitude diagrams (CMDs) and spatial distribution of various stellar populations in our observing field. By focusing on the two stellar streams, we show more about the derivation of their distance and metallicity distributions, and discuss the origin of the streams later in the section. In Section~\ref{sec:conclusion}, we draw discussion and conclusions.

\section{Observation and Data Reduction}\label{sec:data}

\subsection{HSC Observations}

In this study, we use the HSC\footnote{\url{http://www.subarutelescope.org/Observing/Instruments/HSC/index.html}} imager on the 8.2-m Subaru Telescope. HSC consists of 104 $2048\times4096$ science CCDs with a scale of $0\farcs17$ per pixel and covers a total field-of-view (FoV) of 1.5 degree in diameter \citep{Miyazaki12}. In addition, the HSC FoV corresponds to about 190~kpc at the distance of NGC~4631. The HSC observations of the current target were taken during two nights in March 2015 (S15A-046; PI: Tanaka). Our HSC field includes both NGC~4631 and NGC~4656 as shown in Figure~\ref{fig:map_photgrid}. The HSC observations were made with HSC-$g$ and HSC-$i$ filters with the total exposure time of 6.0h and 11.8h, respectively. The weather conditions were excellent with stable seeing of $0\farcs48$ in HSC-$g$ and $0\farcs60$ in HSC-$i$ which are measured on the final stacked images. Table~\ref{tab:obs} summarizes full details of the observations.

To reduce the raw HSC data, we basically adopt the HSC pipeline 3.8.5\footnote{\url{http://hsca.ipmu.jp}}\footnote{\url{http://hsc.mtk.nao.ac.jp/pipedoc\_e/index.html}}, which is based on an earlier version of the LSST pipeline \citep{Axelrod10}, calibrated against SDSS Data Release 7 astrometry. Since the sky-subtraction process in this version of the pipeline did not work well for the case that CCD chips suffer from apparently large objects such as nearby galaxies, we develop and apply our own sky-subtraction method in such a case, as summarized in the Appendix. Since the data were taken in different airmass as shown in Table~\ref{tab:obs}, flux scaling of each CCD chip is carried out before combining the science images (Bosch et al. 2017, in preparation). First, zero-point magnitudes of each CCD chip were estimated based on the stars with SDSS photometry imaged in each chip, Then, the zero-point magnitude of each exposure and the spatial variation over the focal plane, which is expressed as the 5th-order Chebyshev polynomial, are fitted by comparing the fluxes of same star imaged in different exposures. By adopting this flux scaling, the spatial variation of flat fielding and the extinction by airmass are corrected. As a result, the flux is calibrated within the photometric accuracy of 0.034 mag in $g$-band, 0.028 mag in $i$-band. Finally, we obtained a $3\sigma$-clipped mean-stacked image for each band.

\begin{deluxetable*}{cccccc}
\tablecaption{Observation Status for a Target Field \label{tab:obs}}
\tablecolumns{6}
\tablenum{1}
\tablehead{
\colhead{Coordinates} & \colhead{Filter} & \colhead{$t_{\rm exp}$} & \colhead{$N_{\rm exp}$} & \colhead{Airmass} & \colhead{Seeing}\\
\colhead{($\alpha_{2000}/\delta_{2000}$)} & \colhead{} & (s) & \colhead{} & \colhead{} & (arcsec) 
}
\startdata
$12^h42^m59\fs0$ & HSC-$g$ & 21660 & 92 & $1.02-1.60$ & $0\farcs48$ \\
$+32^\circ22\arcmin17\farcs0$ & HSC-$i$ & 42601 & 180 & $1.02-1.92$ & $0\farcs60$ \\
\enddata
\end{deluxetable*}

\subsection{Photometry and Calibration}\label{sec:photcalib}

\begin{figure}[ht!]
\figurenum{1}
\plotone{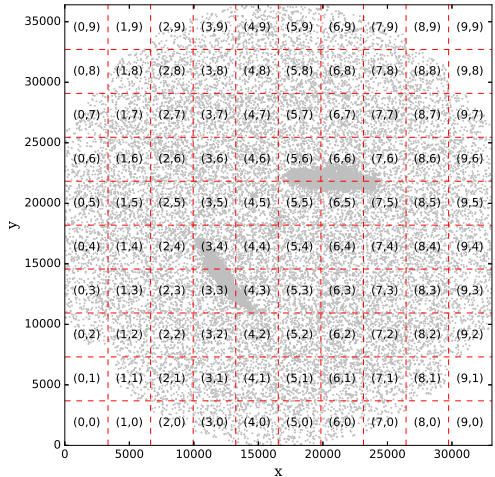}
\caption{The HSC FoV map of subfields divided by $10 \times 10$ red-dashed grids for detection and psf-fitting photometry. Grey points indicate objects from our final photometric catalog selected under the restrictions of PyRAF/DAOPHOT parameters ($merr < 0.5$, $|sharpness| < 0.5$ and $chi < 5$).
\label{fig:map_photgrid}}
\end{figure}

We conducted point-spread function (PSF)-fitting photometry using PyRAF in the similar manner to \cite{Tanaka10}. The PSF models were statistically chosen based on reduced chi-squared computed by the PyRAF/DAOPHOT PSF task. When conducting detection and photometry, we divide a HSC field into several subfields and photometry is carried out for each subfield separately shown by the red dashed-grids of Figure~\ref{fig:map_photgrid} because spatial variations of PSF were caused by a huge FoV of HSC. We detect 2,301,816 (1,594,340) objects in $g$-band ($i$-band). Finally, we merged two independent $g$-band and $i$-band catalogs into a combined catalog (including 1,218,065 objects) using a 3-pixel matching radius.

\begin{figure}[ht!]
\figurenum{2}
\plotone{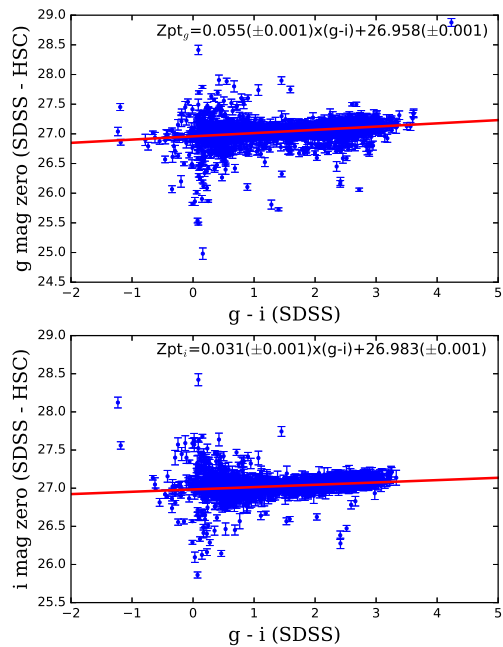}
\caption{Relations between color terms and photometric zero-points of $g$-band ({\it top}) and $i$-band ({\it bottom}) of stellar-like objects selected based on stringent criteria (see text). Red solid lines show least-square regression lines.\label{fig:magzero}}
\end{figure}

We calculated color terms and photometric zero-points per second of both bands by comparing stellar objects from SDSS Data Release 12. The standard objects are selected under the restrictions of SDSS parameters ($psfMagErr < 0.1$) with those from our photometric catalog selected under the restrictions of PyRAF parameters ($merr < 0.01$, $|sharpness| < 0.1$ and $chi < 2$). Figure~\ref{fig:magzero} shows best-fitted relations between color terms and photometric zero-points. For further inspection, we check to what extent photometric zero-points vary with position in our HSC field. The differences between the best-fitted zero-points and ones in the subfields are 0.03~mag in $g$-band and 0.02~mag in $i$-band which are $3\sigma$-clipped standard deviations. This result is consistent with the photometric accuracy of the flux calibrations conducted in the reduction process.

Reddening correction is applied to each object in this HSC field based on the extinction maps of \citet{Schlegel98} and its spatial variation is negligible. In addition, we adopt standard conversion laws, $A_g = 3.793$E(B-V) and $A_i = 2.086$E(B-V), in the SDSS's filter system \citep{Schlafly11}.

\subsection{Completeness}\label{sec:completeness}

\begin{figure*}[ht!]
\figurenum{3}
\plottwo{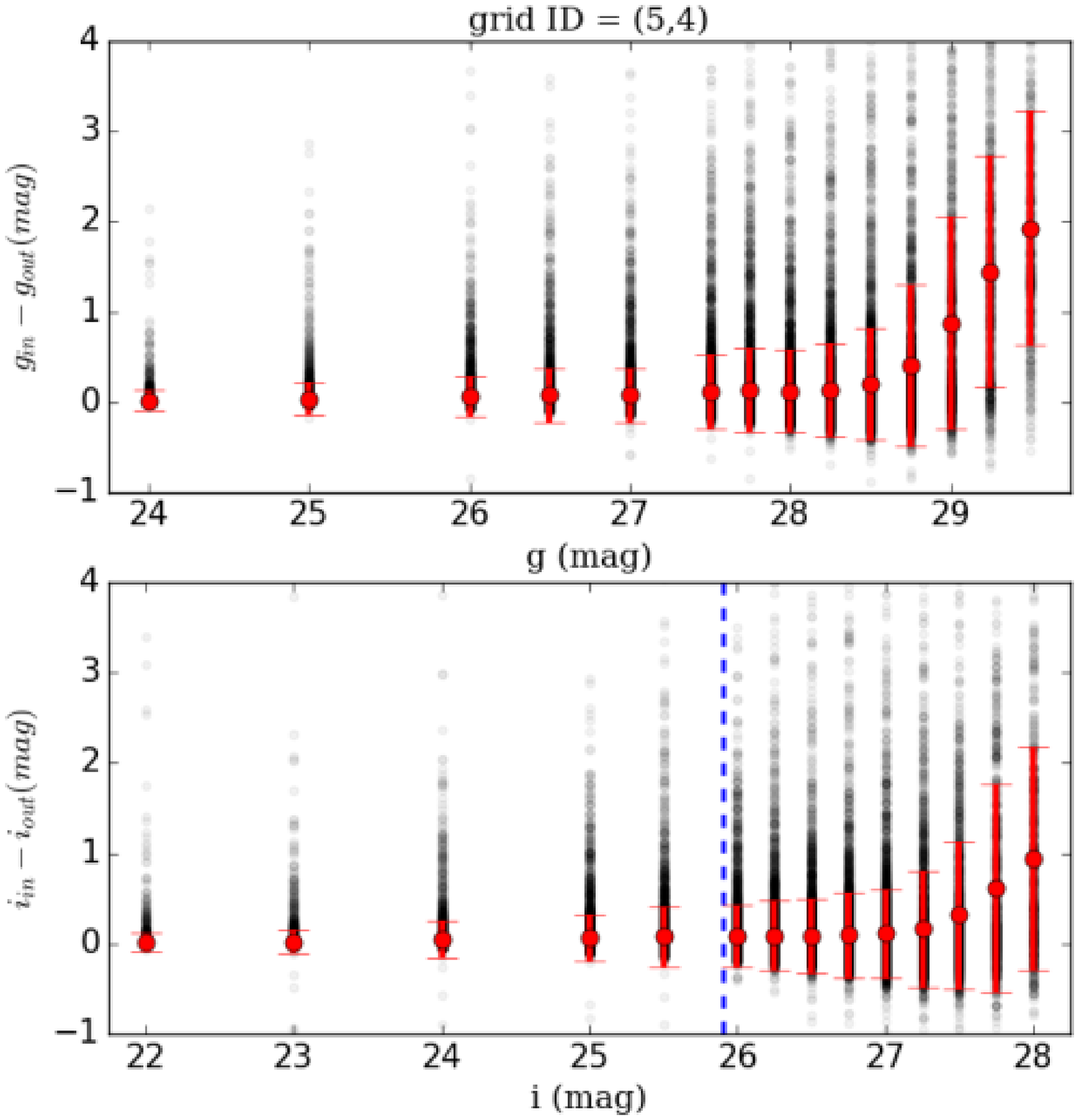}{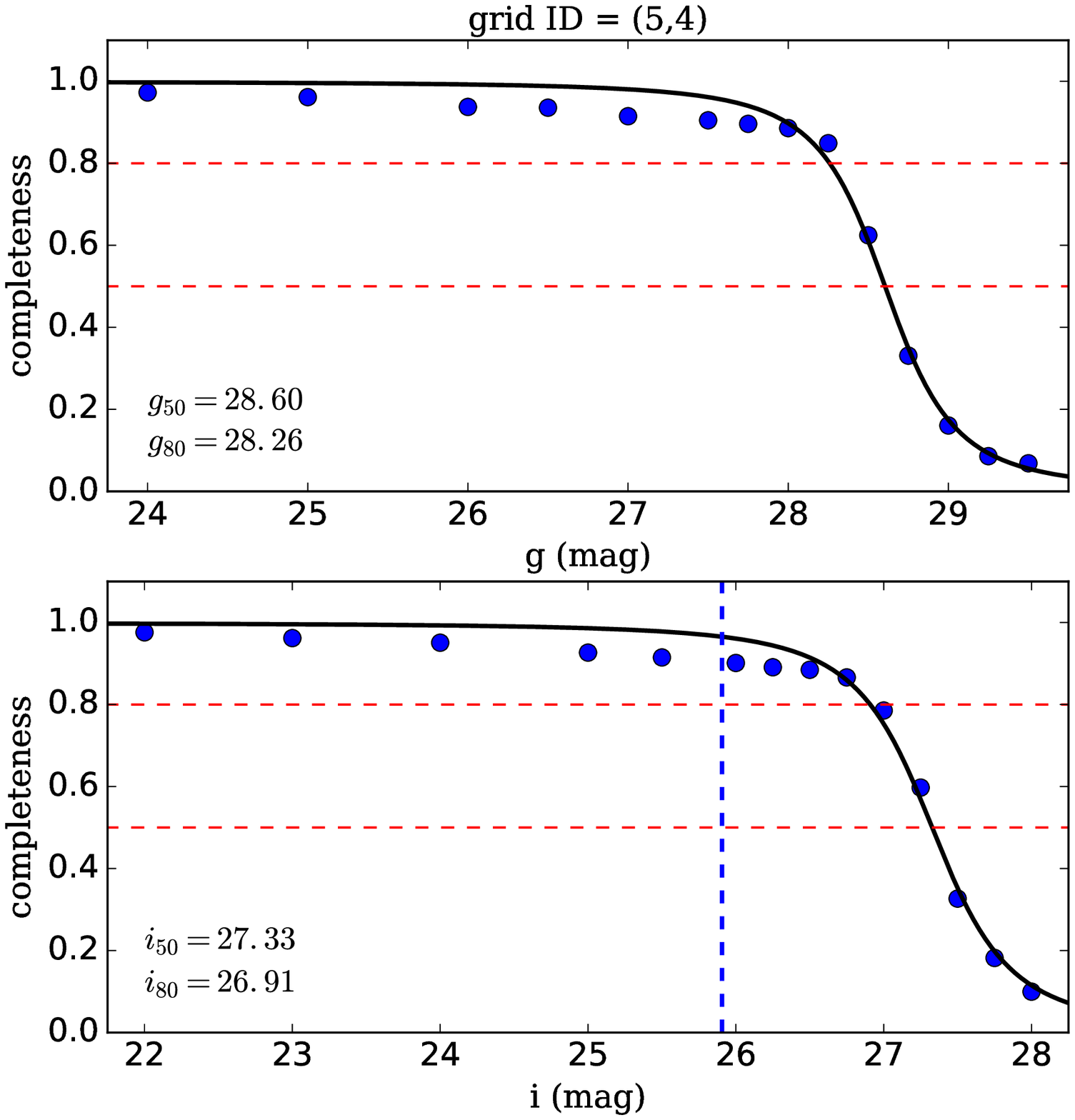}
\caption{{\it Left}: difference between input and output magnitude of artificial stars in the grid number of id $=$ (5,4). Gray-scaled points show the number density of recovered artificial stars, whereas red points and error bars present mean and standard deviation in each magnitude bin. The photometric errors in each grid are calculated from this artificial tests. {\it Right}: estimated completeness as a function of magnitude. The black solid lines of each band show the analytical model described by \cite{Fleming95}. The blue vertical dashed-lines in both $i$-band panels indicate the TRGB magnitude of NGC~4631 \citep{Radburn-Smith11}. 
\label{fig:completefunc}}
\end{figure*}

In order to evaluate incompleteness due to low signal-to-noise ratio and crowding, we have performed the artificial star experiments basically in the similar manner to \citet{Tanaka10}. We have added a sufficient number of artificial stars to each original image in each subfield and each magnitude by using the PyRAF $addstar$ task, and the range of the magnitudes is $24.0 \le g \le 29.5$ and $22.0 \le g \le 28.0$ with a binning step of about 0.25-1.0 mag. To prevent these stars from interfering with one another, we have divided the HSC frames into grids of cells of 100 pixels width and have randomly added one artificial star to each cell for each run. In addition, we constrain each artificial star to have 10 pixels from the edges of the cell (see also \citet{Tanaka10}). About 5,400 stars were configured for each run, and the total number is about 6,274,800 in each image. After detection/photometry were performed in the same manner as in the previous section, we regard an output object as the same one as an input object based on standard deviation of the magnitude difference between the input and output objects within the matching radius. Then, the 50 and 80 percent completeness limits were determined by fitting the analytical model described by \cite{Fleming95}. An example for gird ID $=$ (5,4) is shown in Figure~\ref{fig:completefunc}. 

\begin{figure*}[ht!]
\figurenum{4}
\plotone{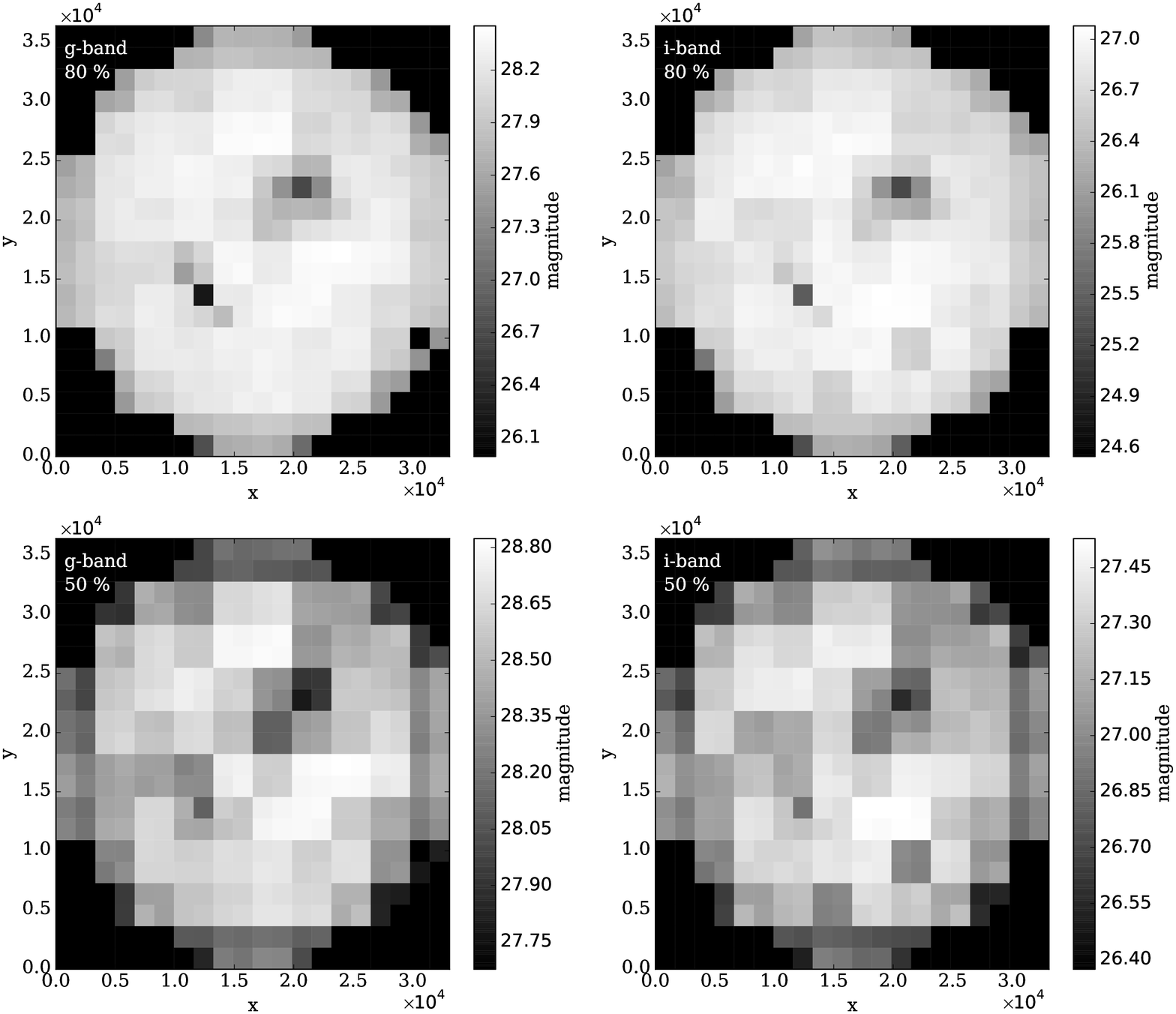}
\caption{Spatial variations of magnitudes with 80\% and 50\% completeness limits. A subfield for detection/photometry was divided into further four sub-subfields in order to draw the spatial variations.\label{fig:completeness}}
\end{figure*}

Figure~\ref{fig:completeness} shows spatial distributions of magnitudes with 80\% and 50\% completeness limits in each band. The completeness limits are somewhat shallow near the edge of the FoV due to low signal-to-noise ratio and close to the two main galaxies in this galaxy group due to crowding. On the other hand, completeness limits of other fields such as NGC~4631's halo are deep enough to reach the tip of red giant branch (TRGB) of the galaxy system.

\section{Results}\label{sec:results}

\subsection{Satellite Galaxies around NGC~4631}\label{sec:dwarf}

\begin{figure*}[ht!]
\figurenum{5}
\plotone{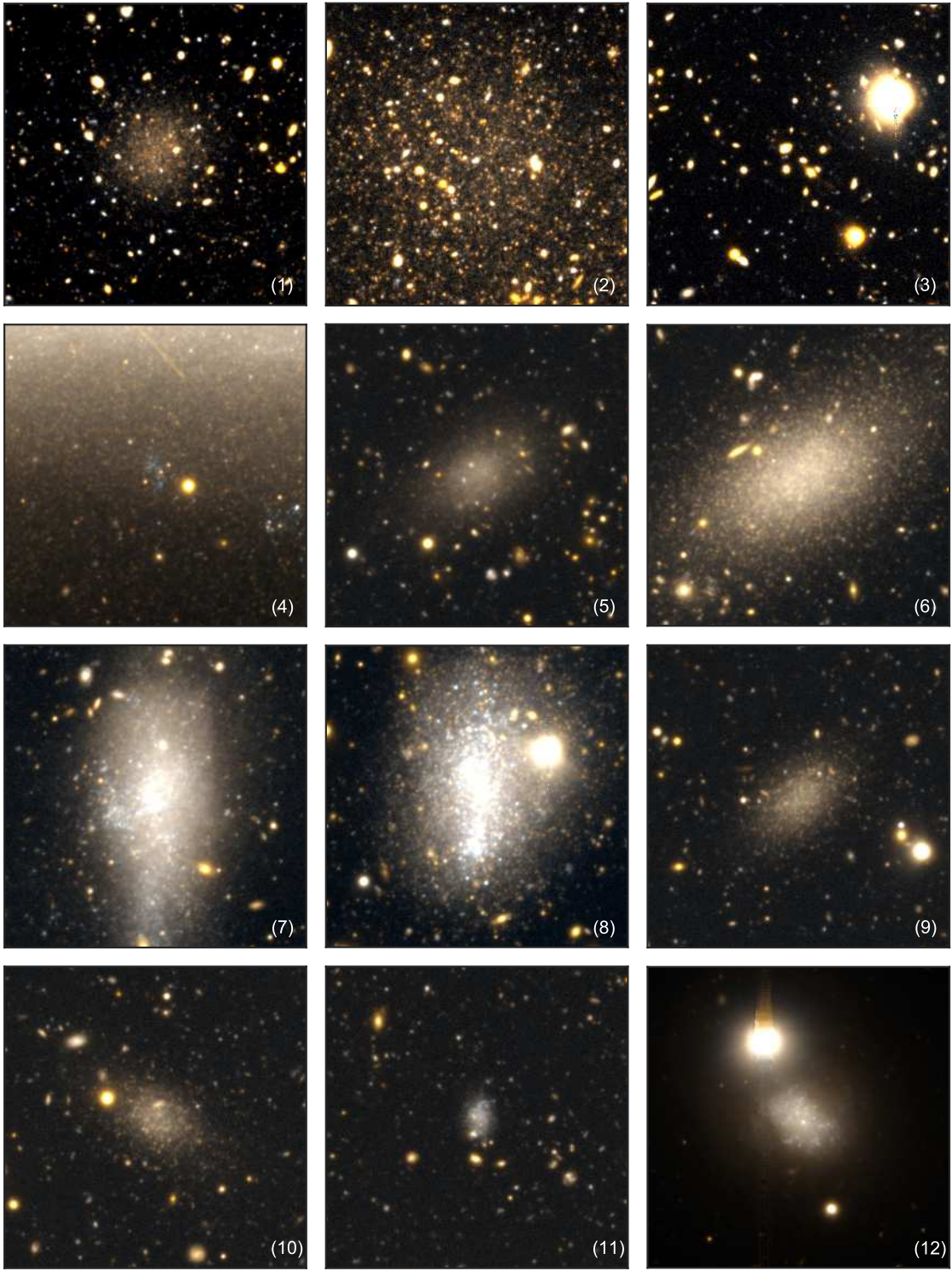}
\caption{Three color composites characterized by HSC-$g$ and HSC-$i$ images. A pseudo image with intermediate color is created from the average image of HSC-$g$ and HSC-$i$ images. Each panel has all the same $0\fdg02 \times 0\fdg02$ FoV, and numbers in each panel are corresponding to ID numbers in Table~\ref{tab:dwarf} and label numbers in Figure~\ref{fig:plot_stream}.
\label{fig:dwarf}}
\end{figure*}

A wide-field, deep and high-resolution HSC image enables us to confirm already-known satellite galaxies and search for still-unknown ones in the galaxy group consisting of NGC~4631 and NGC~4656. Figure~\ref{fig:dwarf} shows three color composites of dwarf candidates, and Table~\ref{tab:dwarf} lists detailed information of their coordinates, stellar populations and corresponding references. Through the visual inspection, we detect 8 new candidate satellite galaxies around NGC~4631 (HSC-5 to 12), and we confirm 3 previously reported ones (HSC-1, 2 and 4). In the composite image at the location of DGSAT-3, a low surface brightness galaxy near NGC~4631 reported by \citet{Javanmardi16}, we do not find a galaxy, and suggest that this was a misclassified object due to blending of foreground stars and background galaxies (see our object HSC-3).

HSC-1 and HSC-11 show somewhat clumpy features located at each of these galaxy systems, but unresolved in our HSC image. The former has a relatively red diffuse structure in Figure~\ref{fig:dwarf}, suggesting that it dominantly consists of an old stellar population, whereas a relatively blue image of the latter implies a young stellar object such as a dwarf irregular galaxy. The other candidates are clearly resolved in this HSC image, for instance, a diffuse low surface brightness galaxy, HSC-2, which is also reported by previous unresolved studies based on small amateur telescopes and HSC-1 \citep{Karachentsev14,Martinez15,Javanmardi16}, has an extended structure consisting of red, old stars as can be seen in the composite image. On the other hand, HSC-4 reported by the HST observation of \citet{Seth05a} is located in the thick disk of NGC~4631 (see also Figure~\ref{fig:plot_stream}), and the existence of the resolved blue stars supports their conclusion that it is a young dwarf galaxy of an age of $\sim30$~Myr estimated from the comparison of their resolved CMD with theoretical isochrones. 

Both HSC-7 and HSC-8 look like dwarf irregular galaxies with blue, young stellar populations, while HSC-12, which is located near a saturated bright foreground star, looks also like a compact dwarf irregular galaxy with blue stars. On the other hand, HSC-5, 6, 9 and 10 are classified as dwarf spheroidal galaxies dominated by red, old stellar populations in appearance. Further details of these candidate dwarf galaxies will be discussed in the forthcoming paper.

\begin{deluxetable*}{ccccc}
\tablecaption{Candidates for Satellite Galaxies \label{tab:dwarf}}
\tablecolumns{5}
\tablenum{2}
\tablewidth{0pt}
\tablehead{
\colhead{ID} & \colhead{RA(J2000)} & \colhead{Dec(J2000)} & \colhead{Stellar Population} & \colhead{Ref.\tablenotemark{\dag}}
}
\startdata
 HSC-1 & $12^h42^m53\fs1$ & $+32^\circ27\arcmin19\farcs0$  & old? (unresolved) & 2,3,4  \\ 
 HSC-2 & $12^h42^m06\fs1$ & $+32^\circ37\arcmin14\farcs8$  & RGB & 2,3,4  \\
 HSC-3 & $12^h41^m08\fs0$ & $+32^\circ26\arcmin50\farcs4$ & not satellite? & 4  \\
 HSC-4 & $12^h41^m50\fs3$ & $+32^\circ31\arcmin01\farcs9$ & MS & 1 \\
 HSC-5 & $12^h43^m44\fs8$ & $+32^\circ32\arcmin02\farcs9$  & RGB & 5 \\
 HSC-6 & $12^h43^m24\fs8$ & $+32^\circ28\arcmin55\farcs3$  & AGB/RGB & 5 \\
 HSC-7 & $12^h40^m10\fs0$ & $+32^\circ39\arcmin31\farcs4$  & MS/RSG/AGB/RGB & 5 \\
 HSC-8 & $12^h41^m47\fs2$ & $+32^\circ51\arcmin24\farcs2$  & MS/RSG/AGB/RGB & 5 \\
 HSC-9 & $12^h40^m53\fs0$ & $+32^\circ16\arcmin55\farcs8$  & AGB/RGB & 5 \\
HSC-10 & $12^h42^m31\fs4$ & $+31^\circ58\arcmin09\farcs3$  & AGB/RGB & 5 \\
HSC-11 & $12^h42^m01\fs7$ & $+32^\circ24\arcmin06\farcs6$  & young? (unresolved) & 5  \\
HSC-12 & $12^h43^m07\fs1$ & $+32^\circ29\arcmin27\farcs3$  & MS/RSG/AGB/RGB & 5  \\
\enddata
\tablecomments{The number of the ID column corresponds to the one of Figure~\ref{fig:dwarf} and Figure~\ref{fig:plot_stream}. The coordinates are central positions of each panel in Figure~\ref{fig:dwarf}. The column of Stellar Population shows which in RGB/AGB/RSG/MS maps each candidate is detected. HSC-1 and HSC-11 are unresolved in our HSC image, while HSC-3, which was previously-reported as a low surface brightness galaxy by a small amateur telescope \citep{Javanmardi16}, is apparently a blending object in our HSC image. \\ \\
\dag \ (1) \citet{Seth05a}, (2) \citet{Karachentsev14}, (3) \citet{Martinez15}, (4) \citet{Javanmardi16}, (5) this work.
}
\end{deluxetable*}

\subsection{Color-Magnitude Diagrams}\label{sec:CMD}

\begin{figure*}[ht!]
\figurenum{6}
\plottwo{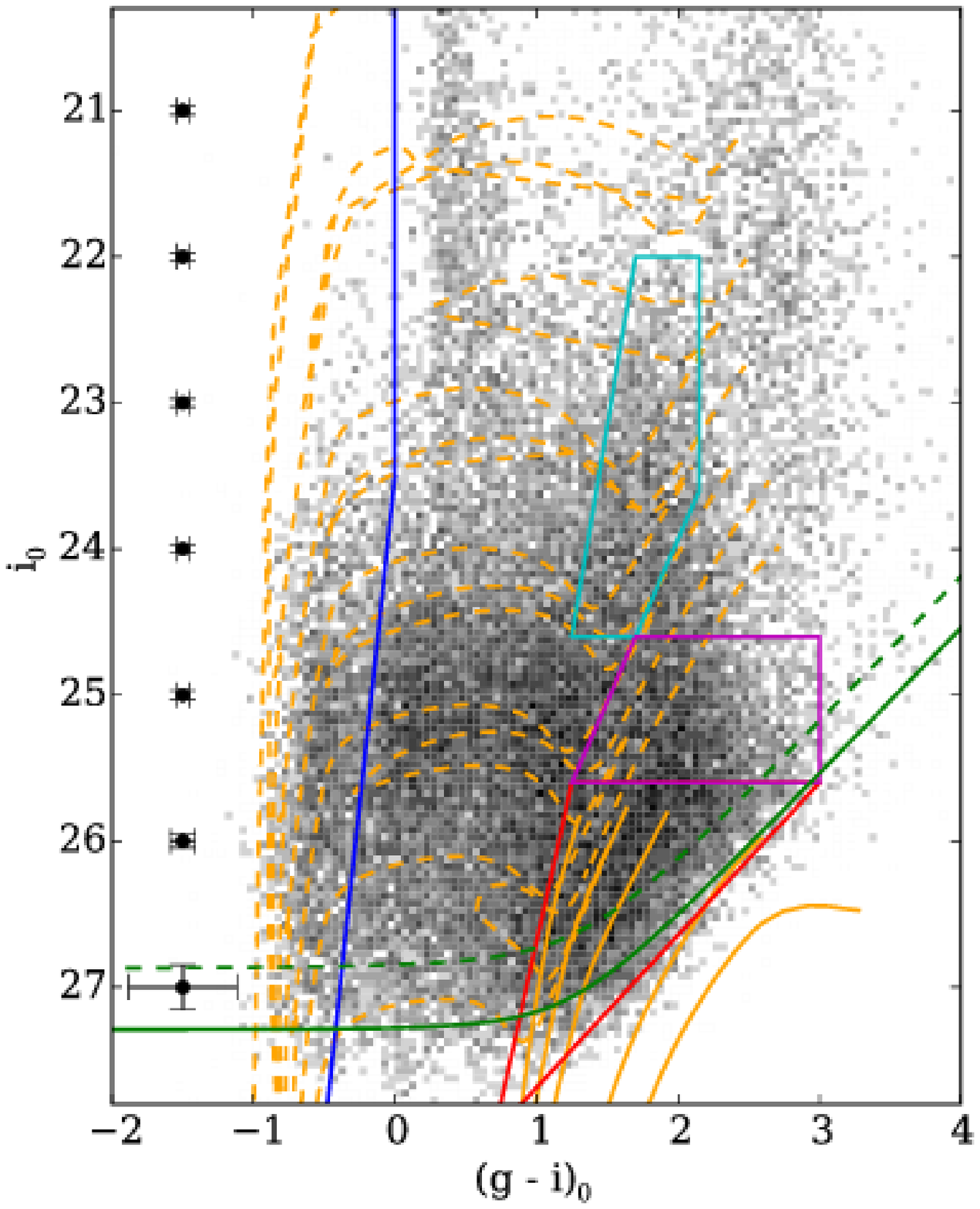}{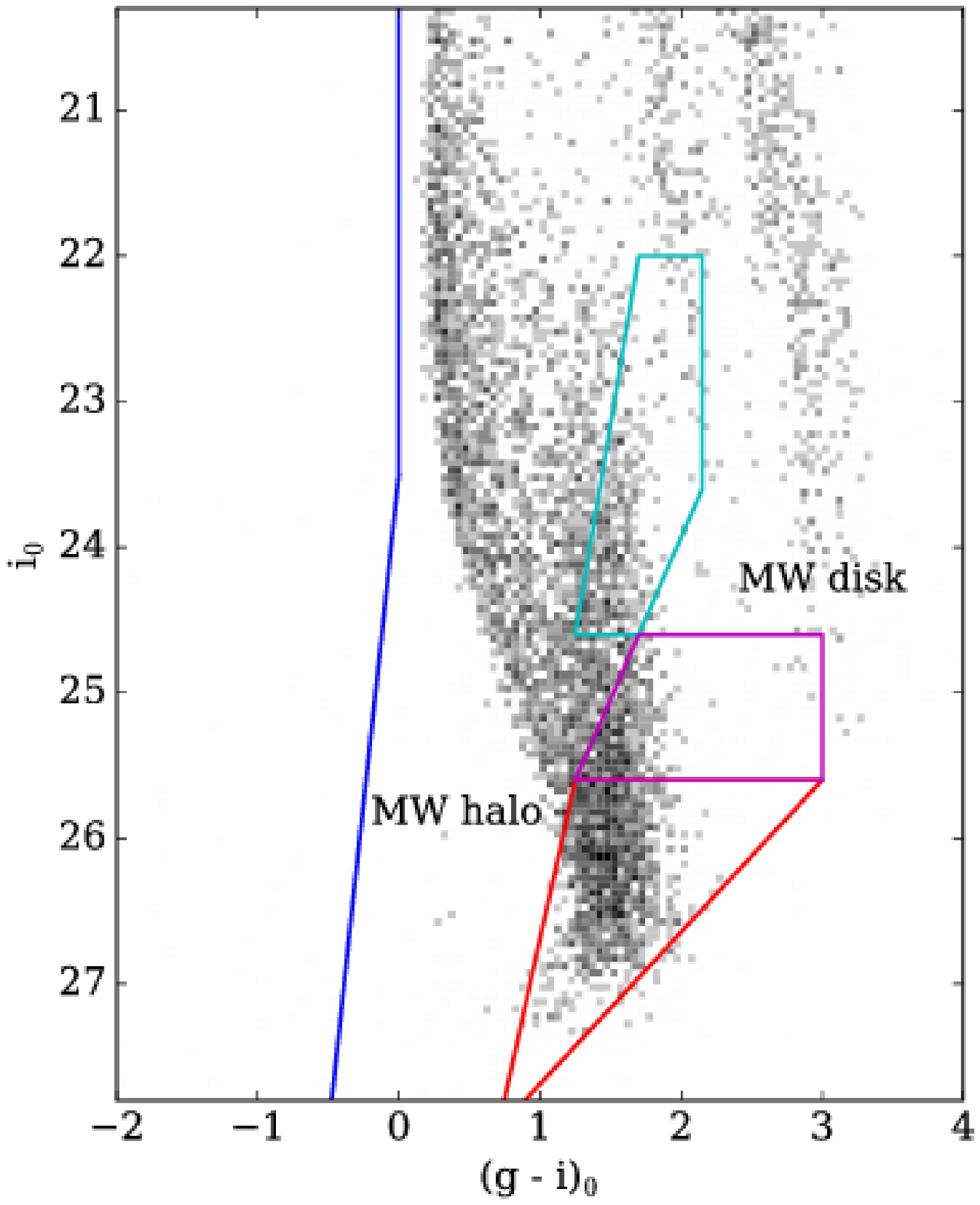}
\caption{{\it Left}: Log-scaled CMD over the entire HSC field including both NGC~4656 and NGC~4631, derived from number densities per $0.05 \times 0.05$ mag box. The orange solid lines are theoretical PARSEC isochrones \citep{Bressan12}, which are scaled to distance of 7.4~Mpc \citep{Radburn-Smith11}, of age 10~Gyr spanning the metallicity range [M/H]$ = -2.28, -1.68, -1.28, -0.68$ and $-0.38$, while the dashed lines are those of [M/H] of $-0.38$ spanning the age range of 4.0, 10.0, 17.8, 31.6, 56.2, 100, 177.8~Myr. The solid and dashed green lines denote the full ranges of the 50\% and 80\% completeness levels of the grid of id $=$ (5,4) of Figure~\ref{fig:map_photgrid}, respectively. The black errorbars correspond to red points, that is, mean values, of the left panels of Figure~\ref{fig:completefunc}. The objects within the red triangle and the magenta and cyan boxes are regarded as RGB, AGB and RSG stars, respectively, while bluer objects than the blue solid line are classified as MS stars. {\it Right}: Log-scaled CMD of Galactic foreground stars simulated from the Besan\c{c}on models \citep{Robin03}, which are convolved with photometric errors and the observational incompleteness effects. The distance interval of the model is set on 0~kpc to 50~kpc. Colored boxes show the same as selection boxes of the left panel.
\label{fig:cmdall}}
\end{figure*}

The left panel of Figure~\ref{fig:cmdall} shows the log-scaled CMD over the entire HSC field derived by plotting number density of stars within $0.05 \times 0.05$ mag boxes. It is compared with PARSEC isochrones \citep{Bressan12} for old population with fixed ages (10~Gyr) and different metallicties ([M/H]$ = -2.28, -1.68, -1.28, -0.68$ and $-0.38$) (orange solid lines) and young population with different age (4.0, 10.0, 17.8, 31.6, 56.2, 100, 177.8~Myr) and fixed metallicity ([M/H] $=-0.38$) (orange dashed lines), assuming that the distance modulus of NGC~4631 is 7.4~Mpc \citep{Radburn-Smith11}. In order to make the CMD, we select stellar-like objects under the restrictions of PyRAF parameters ($merr < 0.5$, $|sharpness| < 0.5$ and $chi < 5$). Apparently, the CMD indicates the existence of the broad red giant branch (RGB) at $(g-i)_0 \sim 1$ and $i_0 \sim 26$ and a young main-sequence (MS) at $(g-i)_0 \sim -1$.

To examine an effect of Galactic foreground contaminations, we produce a synthetic CMD in the same Galactic coordinate and the FoV as this study based on the Besan\c{c}on models \citep{Robin03}, convolving photometric errors and the observational incompleteness (see the right-hand panel of Figure~\ref{fig:cmdall}). The synthetic CMD is consistent with the observational one, and both of them clearly indicate foreground dwarf stars in the MW halo at $(g-i)_0 \sim 0.3$ as well as those in the MW disk at $(g-i)_0 \sim 2.5$ in a bright magnitude range of $ i_0 \lsim 24$. Because NGC~4631 is located at the high Galactic latitude ($b\sim84^\circ$), foreground contaminations of the MW disk are relatively small. In Figure~\ref{fig:cmdall}, the colored boxes show the same criteria of stellar populations as the observed CMD, indicating that a region of MS is little contaminated by foreground stars. In fact, the ratio of the predicted number of foreground stars to the total number of detected objects is about 0.5\%. Although regions of asymptotic giant branch (AGB) and RGB seem largely contaminated, these ratios are actually 1.4\% and 1.0\%, respectively. This ratio for a region of red super giants (RSG) is also small, 2.3\%. Therefore, dominant contaminations in a faint magnitude range of $i_0 \gsim 24.5$ of the observed CMD are high-redshift background galaxies which are unresolved point sources in ground-based telescopes. 
In spite of the fact that analysis based on selection boxes of stellar populations on CMDs is expected to largely reduce effects from background galaxies, it is not easy to completely distinguish local signals of substructures from heavy contaminations due to local density peaks originated from cosmic variance. We thus confine ourselves to consider the local overdensities, which are evident from the visual inspection of the color composite image (Figure~\ref{fig:cmdall}) in this study or already suggested in some previous studies. On the other hand, regarding statistical estimation of physical properties of spatially-spread structures in Section~\ref{sec:distance} and \ref{sec:MDF}, we note that the usage of control fields shown in the next section is effective, provided that background galaxies are uniformly distributed in these fields.

\subsection{Spatial Density Distributions}\label{sec:substructure}

\begin{figure*}[ht!]
\figurenum{7}
\plotone{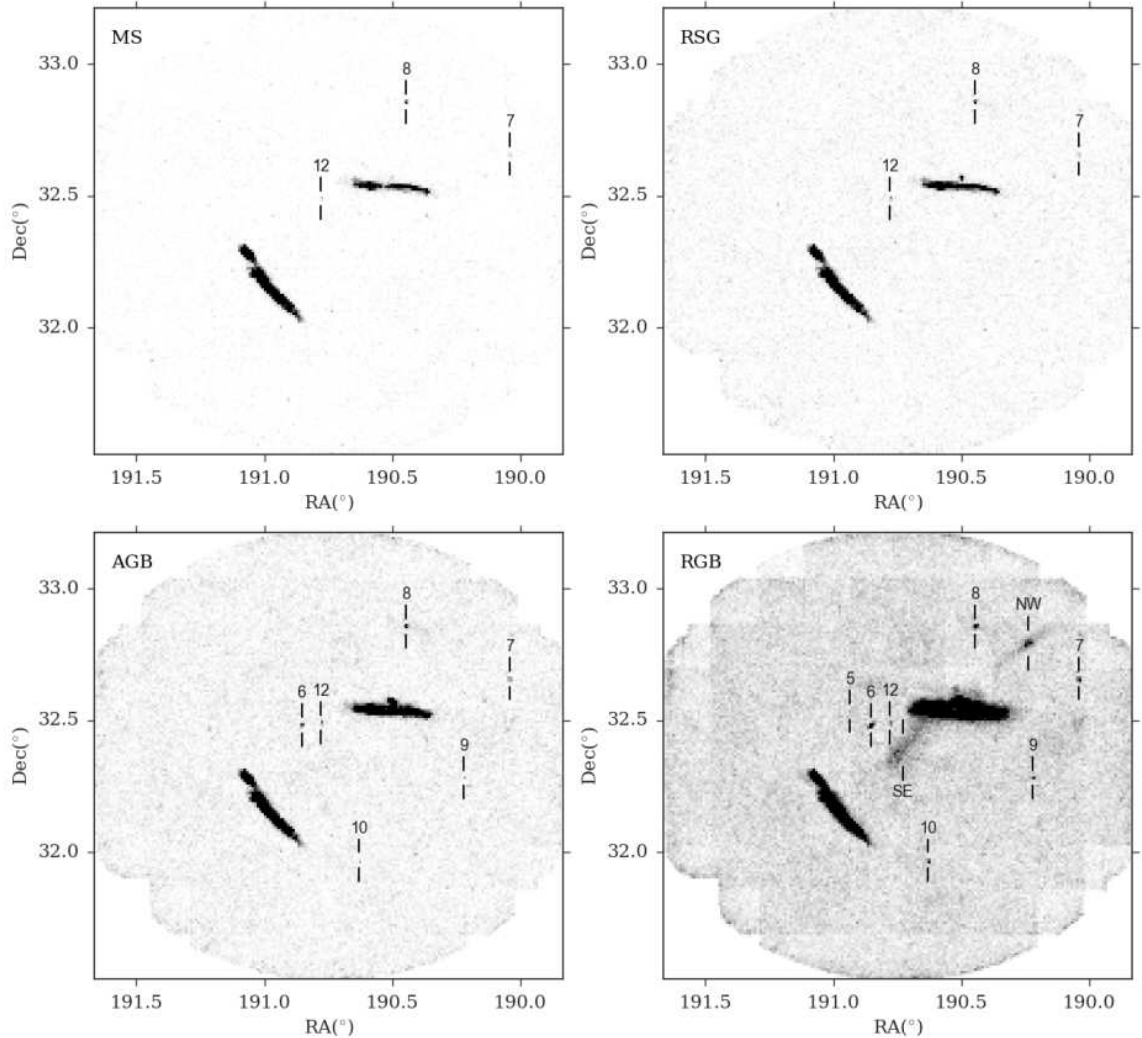}
\caption{Spatial density distributions characterized by MS, RSG, AGB and RGB stars shown in Figure~\ref{fig:cmdall}. This density means amounts of photometric objects per 0.01 $\times$ 0.01 square degrees corresponding to 74 $\times$ 74~kpc$^2$. Resolved candidates for dwarf galaxies newly found in this study are labeled to correspond with Figure~\ref{fig:dwarf} and Table~\ref{tab:dwarf}. The RGB map ({\it bottom-right}) clearly indicates the two resolved old stellar streams at the north-west side (Stream~NW labeled as "NW") and the south-east side (Stream~SE labeled as "SE") of NGC~4631 as well as some as-yet-unknown candidates of dwarf galaxies around NGC~4631 and vertically extended disk of the galaxy, although there are apparently artificial patchy structures due to faint magnitudes near the completeness limits and crowding.\label{fig:mapall}}
\end{figure*}

We divide our photometric objects into the four groups, which are characterized by MS, AGB, RSG and RGB, based on the color cuts shown in the CMD. However, we limit to brighter objects than 50\% completeness limits in selections of MS and RGB. In Figure~\ref{fig:mapall}, we display the spatial distribution for the number density of photometric objects per 0.01 $\times$ 0.01 square degrees corresponding to 74 $\times$ 74~kpc$^2$, for each of the CMD areas enclosed with different color lines in Figure~\ref{fig:cmdall}. These maps indicate many interesting substructures characterized by previously known/unknown dwarf companion galaxies and stellar streams around NGC~4631. In particular, some of dwarf candidates discussed in section~\ref{sec:dwarf} are identified as overdensity regions in these maps. For example, the spatial density map of MS stars shows a remarkable overdensity region, HSC-8, which is apparently a star-forming dwarf galaxy, at the north part of NGC~4631 as well as main bodies of NGC~4631 and NGC~4656. The spatial density map of RSG stars clearly shows a core of NGC~4627, a dwarf Elliptical (dE) satellite of NGC~4631, in addition to the objects presented in the MS map. In the AGB map, other several dwarf satellites with intermediate age populations such as HSC-6 exist between NGC~4631 and NGC~4656.

The RGB map clearly indicates the two resolved stellar streams at the north-west side (Stream~NW) and the south-east side (Stream~SE) of NGC~4631 previously reported by \citet{Martinez15}, although there are apparently artificial patchy structures due to faint magnitudes near the completeness limits and crowding. Both streams mainly consist of likely old, possibly 10~Gyr or even older stellar populations because we cannot detect any counterparts to the two streams in the other maps although AGB stars are slightly concentrated in the core of Stream~NW. Furthermore, the lack of young stellar populations in Stream~SE and NW is consistent with spatial consistent with lack of spatial correlation between these streams and the H$_{\rm I}$ distribution discussed in \citet{Martinez15}. In the subsequent sections, we investigate the photometric properties of these two stellar streams based on the resolved stellar population.

\begin{figure}[ht!]
\figurenum{8}
\plotone{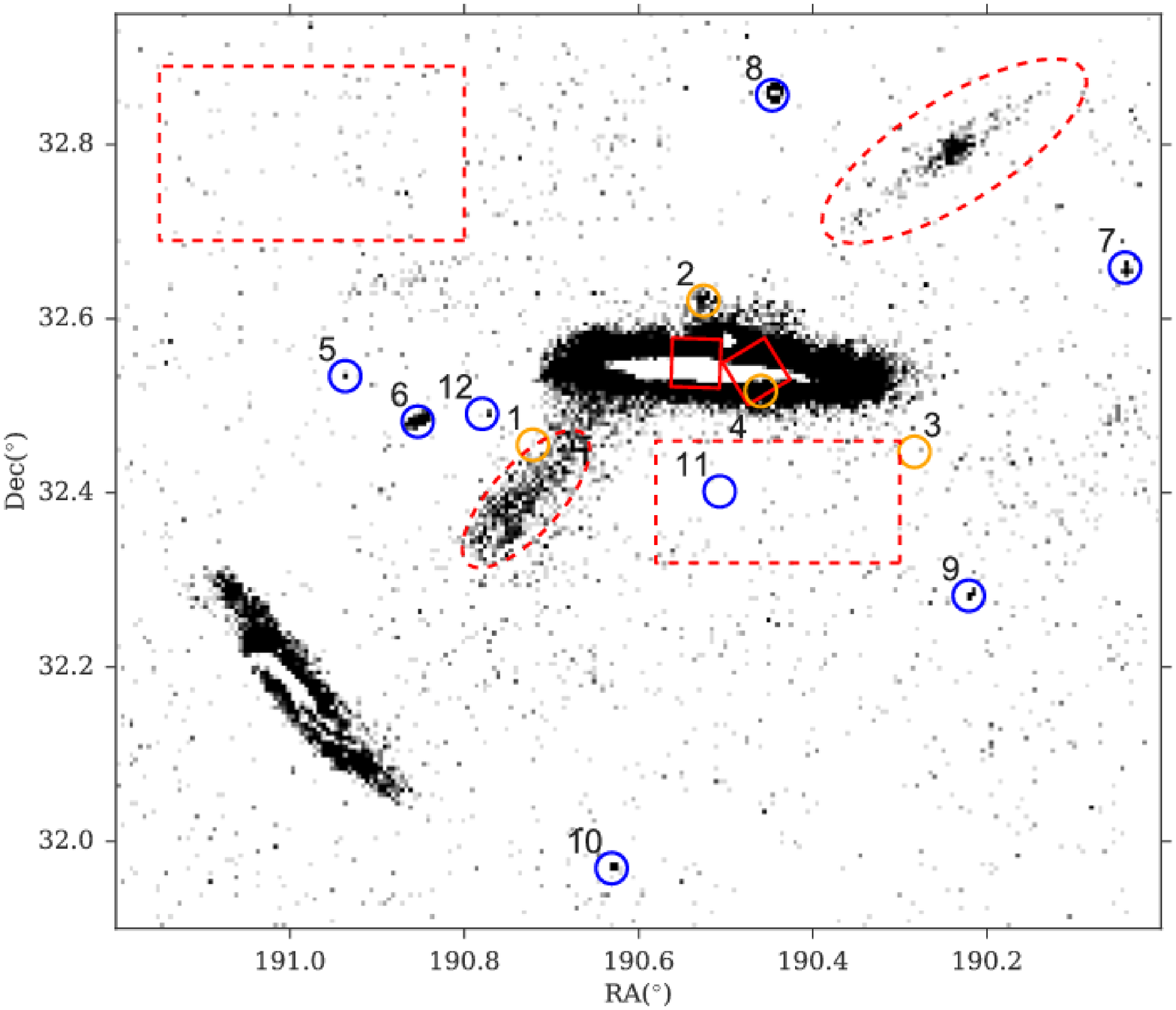}
\caption{A zoomed spatial density distribution of RGB stars. In order to feature substructures around NGC~4631, only pixels with higher S/N than 3 are characterized by gray-scale. The two red dashed-ellipses indicate Stream~SE and NW, while the two red dashed-rectangles illustrate control fields for statistically estimating background and foreground contaminations in these stream fields. The four orange circles present the candidates of dwarf galaxies reported by the previous studies \citep{Seth05a,Karachentsev14,Martinez15,Javanmardi16}. The eight blue ones show the candidates of dwarf galaxies newly detected through the visual inspection in this study. In addition, the two red solid rectangles illustrate the GHOSTS HST fields \citep{Radburn-Smith11}.
\label{fig:plot_stream}}
\end{figure}

The star count maps whose noises are reduced also exhibit some as-yet-unknown candidates of dwarf galaxies around NGC~4631 as well as the stellar streams and the vertically extended disk of the galaxy. Figure~\ref{fig:plot_stream} shows the zoomed spatial density distribution of RGB stars (displayed in the bottom-right panel of Figure~\ref{fig:mapall}) to highlight substructures around NGC~4631. The density is estimated by counting the number of photometric objects within a bin of 0.005 $\times$ 0.005 square degrees corresponding to 37 $\times$ 37~kpc$^2$, and overdensities with a signal over an $S/N$ of 3 are marked in the map, assuming that the global background noise are distributed in Poisson statistics. The two red dashed-ellipses in the map indicate Stream~SE and NW, while the two red dashed-rectangles illustrate control fields for statistically estimating background and foreground contaminations in these stream fields (see also the next section). Therefore, these control fields are located at almost the same projected vertical distance from NGC~4631's disk as the stream fields, and does not have any remarkable substructures of high $S/N$ within these fields. The four orange circles present the candidates of dwarf galaxies reported by the previous studies \citep{Seth05a,Karachentsev14,Martinez15,Javanmardi16}. The eight blue ones show candidate new dwarf galaxies identified through the visual inspection in this study (see Section~\ref{sec:dwarf}). We note that one of these candidates is located in the control field for Stream~SE, but our analysis based on the resolved stellar populations is not affected because this is unresolved. Besides, the other dwarf candidates without high $S/N$ in the figure indicate unresolved objects such as HSC-1 and HSC-11. The reason that the signal of HSC-12 is apparently deflected to the right side from the center is due to a saturated bright star overlapping substantially with the dwarf galaxy. In addition, since HSC-4 located at the old thick disk is a young dwarf galaxy detected by the HST observation \citep{Seth05a}, it mainly consists of MS stars. The resolved stellar populations based on which the new dwarf candidates are visually identified are summarized in Table~\ref{tab:dwarf}.

\subsection{TRGB distance to the streams}\label{sec:distance}

The TRGB, which is driven by core helium ignition, is a useful indicator to estimate the distance to resolved old stellar systems such as nearby galaxies as well as globular clusters \citep[e.g.,][]{Salaris05,Tanaka10,Tanaka11}. 

The $i$-band magnitude of the TRGB in metal-poor populations of [Fe/H]$ \lsim -0.7$ dex changes by less than 0.1 mag \citep{Lee93}. Therefore, the TRGB was historically detected as a sharp cutoff of the $i$-band luminosity function (LF) with the application of an edge-detection algorithm such as Sobel filter \citep[e.g.,][]{Madore95,Sakai96}. However, the edge detection algorithm was confronted with the high levels of Poisson noise that abound in the poorly populated structures of galaxy halos. On the other hand, the maximum-likelihood TRGB detection method, which a predefined model LF is fitted to the observed distribution of the stars, is robust against the strong Poisson noise \citep[e.g.,][]{Mendez02,Makarov06}. In a more recent variation on the maximum-likelihood method, a Bayesian inferential approach for the TRGB detection has been developed due to dramatic computational improvement \citep[e.g.,][]{Conn11,Conn12,Tollerud16}. The traditional maximum-likelihood approach is based on the uncertain assumption that an estimate is normally distributed, whereby a Bayesian approach can estimate a full picture of a complex probability distribution even in a poorly populated structure.

\begin{figure*}[ht!]
\figurenum{9}
\plotone{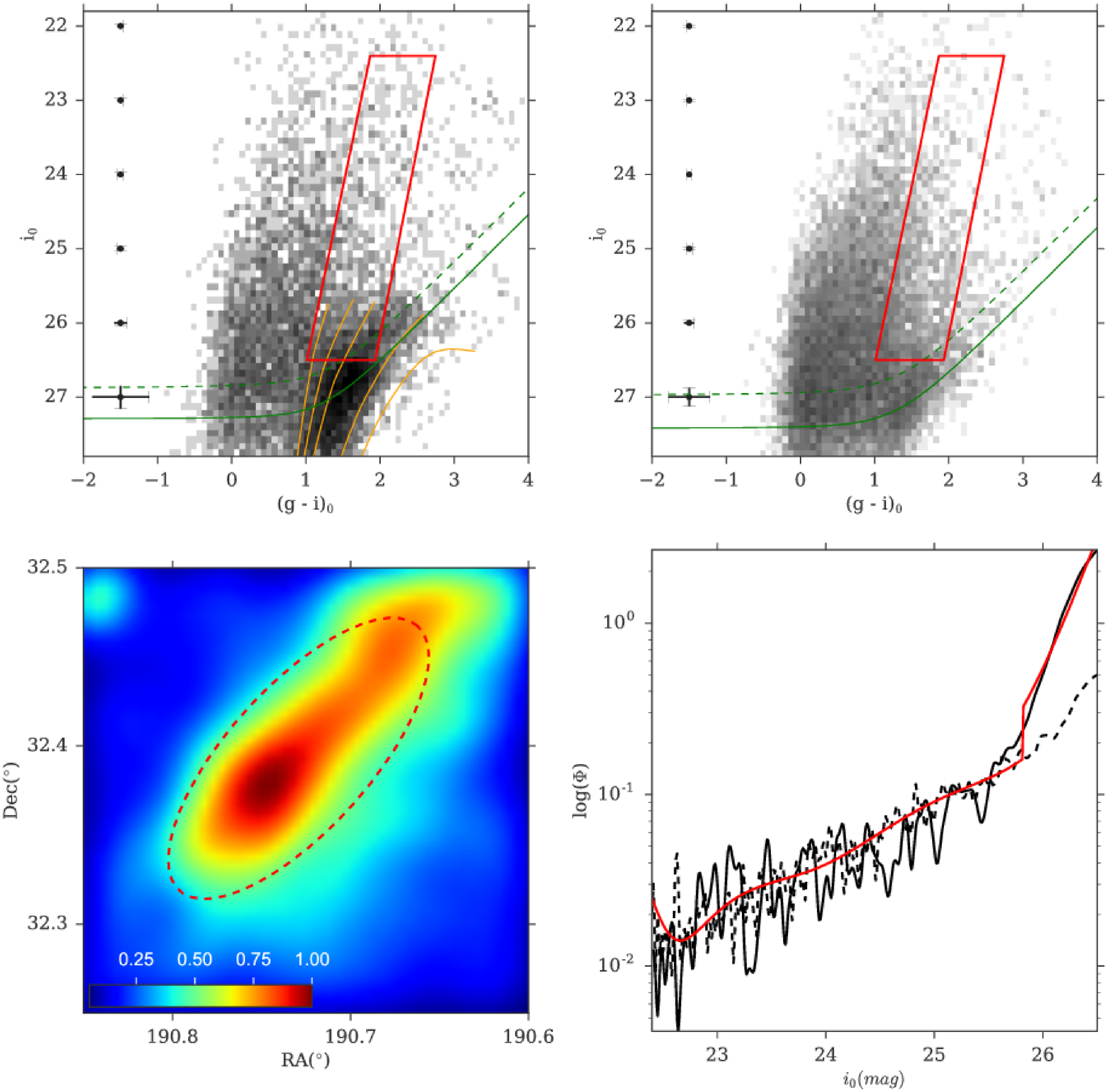}
\caption{{\it Top-Left}: a log-scaled CMD of Stream~SE, which is normalized by area of a bottom red dashed ellipse in Figure~\ref{fig:plot_stream}.  {\it Top-Right}: a log-scaled CMD for its control field, which is normalized by area of a bottom red dashed rectangle in Figure~\ref{fig:plot_stream}. {\it Bottom-Left}: KDE which is the probability distribution function of individual object positions transformed into a smooth surface density function \citep{Silverman86}. The red dashed ellipse corresponds to the one of Figure~\ref{fig:plot_stream}. {\it Bottom-Right}: the logarithmic LF for all stars within the red rectangle of the CMD. The black solid line shows a normalized LF of the stream field, while the black dashed line indicates a normalized LF of the control field. Furthermore, the red solid line presents the best-fit LFs based on EAP estimation (see also Table~\ref{tab:mcmcresult}).\label{fig:stream_SE}}
\end{figure*}

\begin{figure*}[ht!]
\figurenum{10}
\plotone{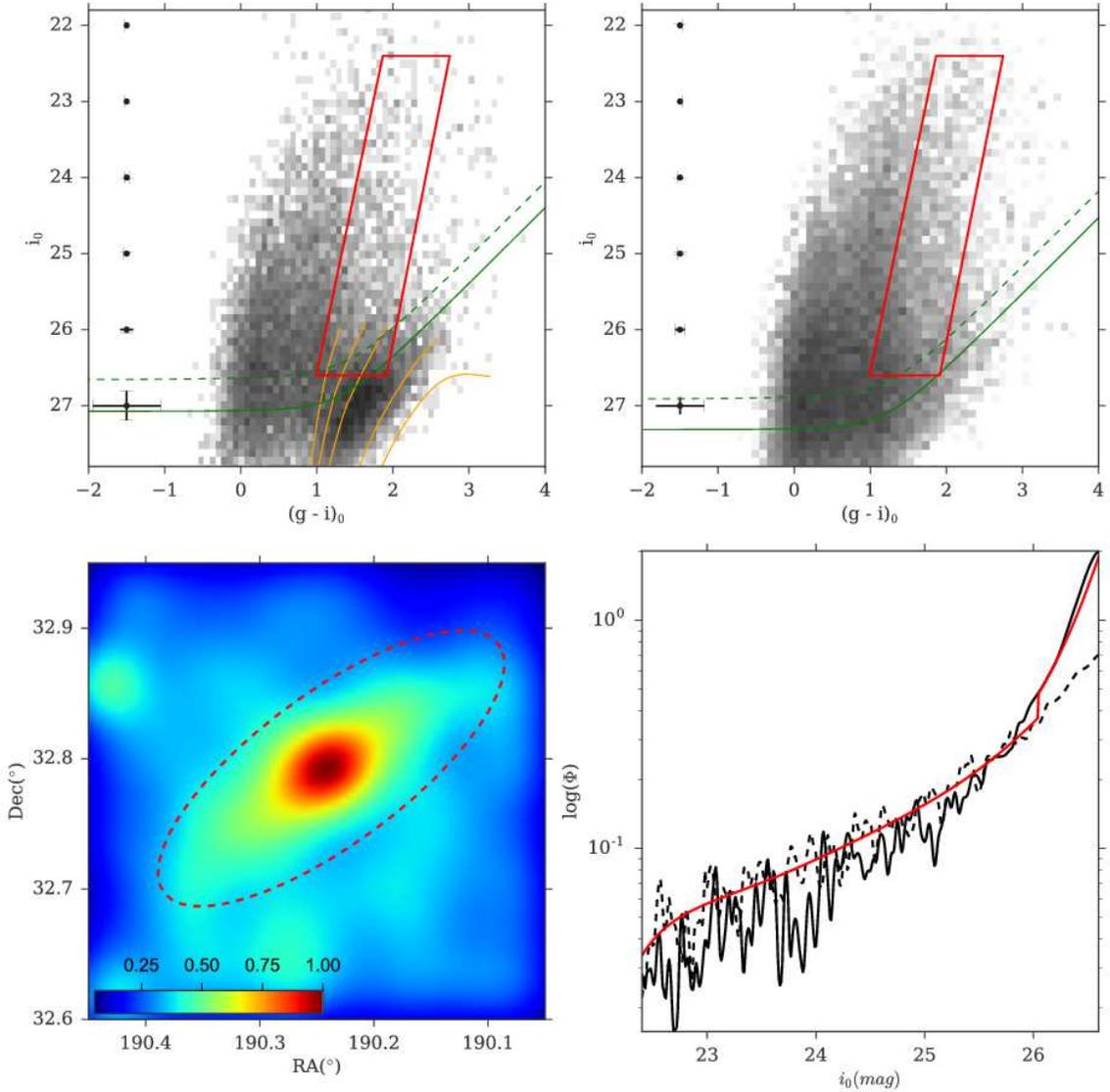}
\caption{Same as Figure~\ref{fig:stream_SE}, but for Stream~NW.\label{fig:stream_NW}}
\end{figure*}

Therefore, we adopt the Bayesian method as described in \citet{Conn11}, to derive the distance to the two streams around NGC~4631. The plots for Stream~SE and NW are summarized in Figure~\ref{fig:stream_SE} and \ref{fig:stream_NW}, respectively. The top-left and top-right panels of these figures show log-scaled CMDs of the stream and its control field, respectively. Using all detected objects, within the red rectangles of these CMDs, we construct completeness-corrected LFs smoothed with a Gaussian kernel, assuming that each object normally distributes at the measured $i$-band magnitude, $\mathcal{N}(m,\sigma)$, where $m$ and $\sigma$ are the $i$-band magnitude and its photometric error calculated from artificial star experiments in Section~\ref{sec:completeness}, respectively \citep{Sakai96}. These LFs are shown in the bottom-right panels of Figure~\ref{fig:stream_SE} and \ref{fig:stream_NW}. 

The background contaminations are not negligible in the magnitude range of faint RGBs in this study. Therefore, we reduce the effect based on the matched filter method adopted in \citet{Conn12}. To do so, we introduce a kernel density estimation (KDE), $\mathcal{W}(\alpha,\delta)$, which is the probability distribution function of individual object positions transformed into a smooth surface density function with a Gaussian kernel and a bandwidth $h$ that is tuned to each distribution with the prescription $h=1.06\sigma N^{-0.2}_{\ast}$ \citep[][where $\sigma$ is the standard deviation of the samples and $N_{\ast}$ is the total number of samples.]{Silverman86}, instead of radial density profiles of dwarf galaxies of \citet{Conn12}. Then, the LF, $\Phi(m)$, within a particular magnitude bin, $m$, is given as
\begin{equation}
\Phi(m) = \sum^{ndata}_{k=1} \mathcal{N}(m_k,\sigma_k)(m)\mathcal{C}_k\mathcal{W}(\alpha_k,\delta_k),
\label{eq:lfobs}
\end{equation}
where $\sigma_k$, $\mathcal{C}_k$ and $\mathcal{W}(\alpha_k,\delta_k)$ are the photometric error, the inverse of completeness and the KDE weight of $k$th object, respectively, and $ndata$ is total number of objects within the red parallelograms of the CMDs. Using resampling data of extinction-corrected $i$-band magnitude, $\{m_1, \cdots,m_N\}$, derived from Equation~(\ref{eq:lfobs}), we calculate the following likelihood function,
\begin{equation}
\mathcal{L}(\{m_1, \cdots,m_N\}|m_{\rm TRGB},a)=\prod_{n=1}^{N}\phi(m_n|m_{\rm TRGB},a),
\end{equation}
where the model LF, $\phi(m_n|m_{\rm TRGB},a)$, is described by
\begin{eqnarray}
\left\{ \begin{array}{ll}
\phi_{\rm RGB}(m_n)+\phi_{\rm BG}(m_n),\ (m_n \geq m_{\rm TRGB})\\
\phi_{\rm BG}(m_n),\ (m_n < m_{\rm TRGB})
\end{array}\right.
\label{eq:lfmodel}
\end{eqnarray}
where $\phi_{\rm RGB}(m_n)$ is $10^{a(m-m_{\rm TRGB})}$ \citep{Makarov06} and $\phi_{\rm BG}(m_n)$ is 6th-order polynomial. Then, we manually determine the fraction of background objects, $f=D_{\rm BG}/D_{\rm SIGNAL}$, by calculating the average density of objects in the control field $D_{\rm BG}$ and in the stream field $D_{\rm SIGNAL}$ in order to match the observational LF from Equation~(\ref{eq:lfobs}) with the model LF from Equation~(\ref{eq:lfmodel}), namely, 
\begin{eqnarray}
\int_{m_{\rm TRGB}}^{m_2}\phi_{\rm RGB}(m_n) dm &=&1-f\\
\int_{m_1}^{m_2}\phi_{\rm BG}(m_n) dm &=& f,
\end{eqnarray}
based on \citet{Conn11}. Eventually, a posterior distribution function is described by 
\begin{eqnarray}
&p&(m_{\rm TRGB},a|\{m_1, \cdots,m_N\}) \nonumber \\
&\propto& \mathcal{L}(\{m_1, \cdots,m_N\}|m_{\rm TRGB},a)p(m_{\rm TRGB})p(a),
\end{eqnarray}
where $p(m_{\rm TRGB})$ and $p(a)$ are the simplest non-informative prior distributions, that is, $uniform$, due to the principle of insufficient reason. The parameters, $m_{\rm TRGB}$ and $a$, are currently chosen for the model by a Markov Chain Monte Carlo (MCMC) algorithm. For sampling of the parameters and marginal likelihood estimation, we apply the standard Metropolis-Hasting algorithm with the symmetric proposal distribution, that is, $Gaussian$. Our algorithm based on 32 MCMC chains and 101,000 iterations, which the first 1,000 steps are discorded as burn-in, each chain adequately converge beyond burn-in, and has totally over 3 million effective random numbers in each stream.

\begin{figure*}[ht!]
\figurenum{11}
\plottwo{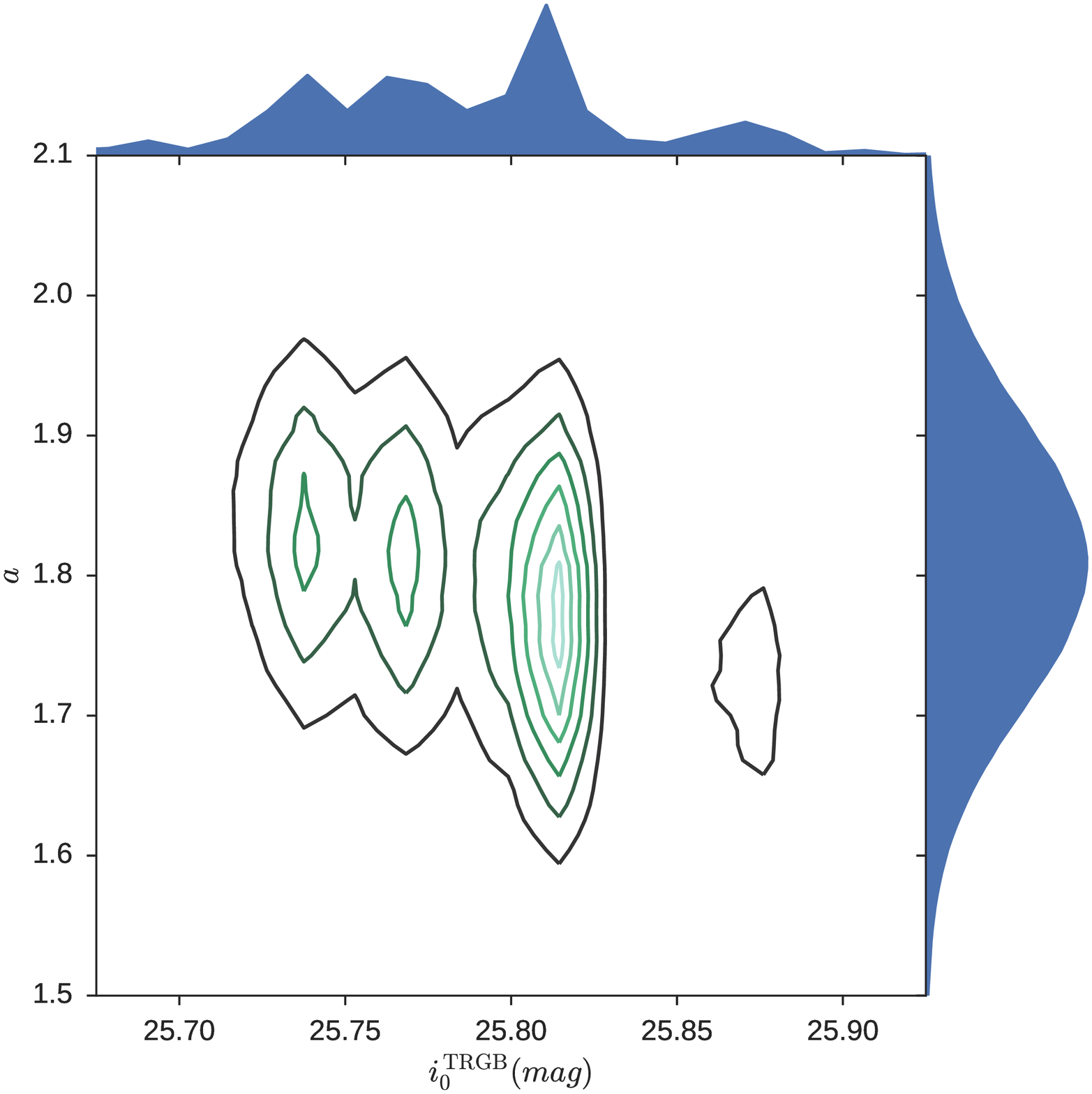}{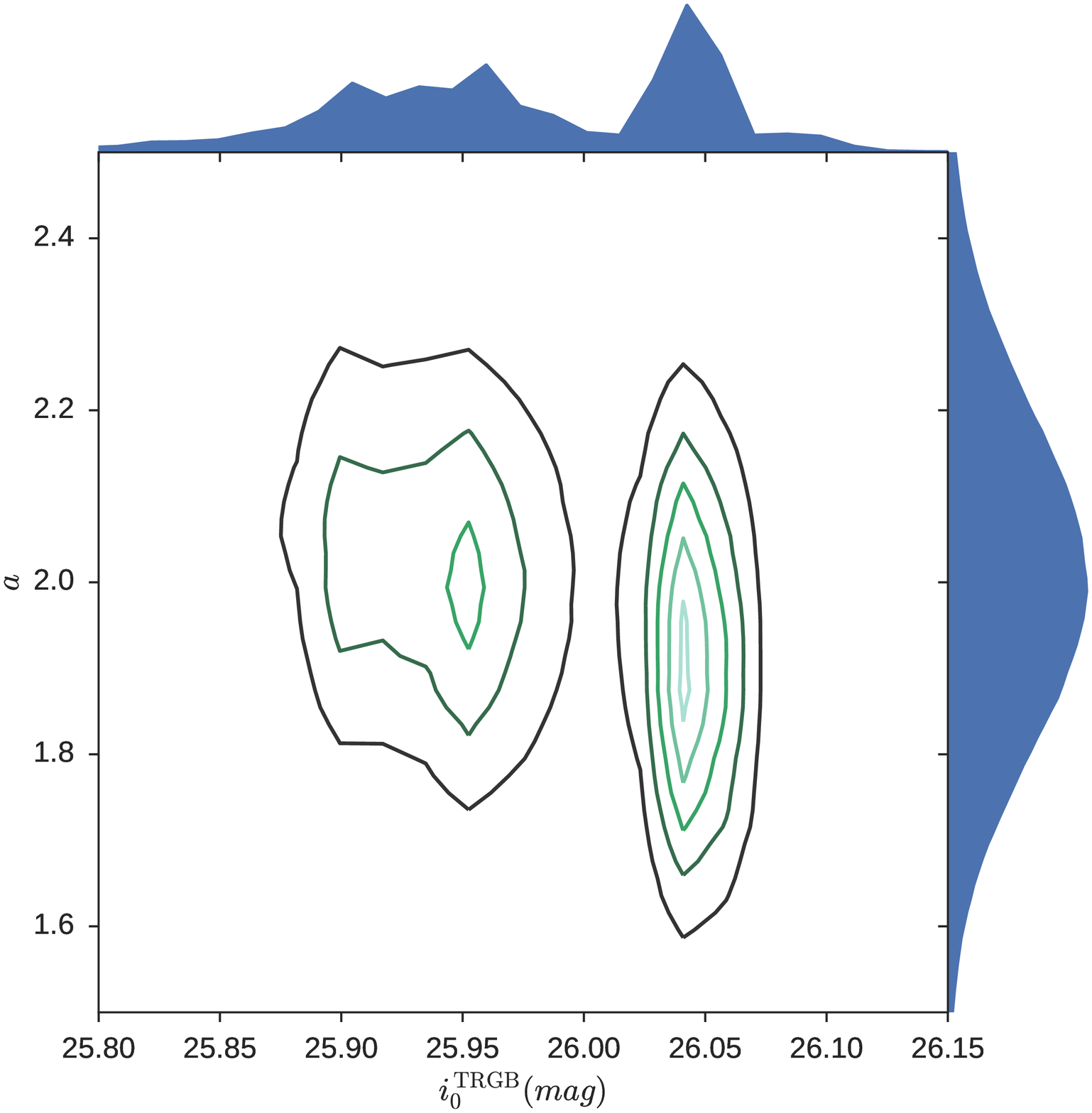}
\caption{Posterior distributions and contour maps for the extinction-corrected TRGB magnitude and the LF slope $a$ of Stream~SE ({\it left}) and NW ({\it right}) based on the effective random numbers generated from our MCMC algorithm.\label{fig:mcmcresult}}
\end{figure*}

Figure~\ref{fig:mcmcresult} shows posterior distributions and contour maps for the extinction-corrected TRGB magnitude and the LF slope $a$ of Stream~SE and NW based on the effective random numbers. The finally-estimated distance to the two streams are summarized in Table~\ref{tab:mcmcresult}, assuming that the absolute TRGB magnitude is $M^{\rm TRGB}_{i,{\rm SDSS}} = -3.44 \pm 0.10$ \citep{Bellazzini08}, suggesting that Stream~NW is relatively more distant from us than Stream~SE. The best-fit LFs based on Expected a posteriori (EAP) estimation are shown with red solid lines in the bottom-right panels of Figure~\ref{fig:stream_SE} and \ref{fig:stream_NW}. 

\begin{deluxetable*}{c|ccccccc}
\tablecaption{Distance to the Streams \label{tab:mcmcresult}}
\tablenum{3}
\tablewidth{0pt}
\tablehead{
\colhead{$(Mpc)$} & \colhead{MAP\tablenotemark{a}} & \colhead{EAP\tablenotemark{b}} & \colhead{5\%} &  \colhead{25\%} & \colhead{50\%} & \colhead{75\%} & \colhead{95\%}
}
\startdata
Stream~SE & 6.90 & 7.10 & 6.22 & 6.84 & 6.97 & 7.09 & 7.29\\
Stream~NW & 7.48 & 7.91 & 6.44 & 7.41 & 7.59 & 7.87 & 7.97
\enddata
\tablenotetext{a}{Maximum a posteriori}
\tablenotetext{b}{Expected a posteriori}
\tablecomments{The 90\% (50\%) credible intervals for the distance to Stream~SE and NW are [6.22, 7.29] ([6.84, 7.09]) and [6.44, 7.97] ([7.41, 7.87]) Mpc, respectively.}
\end{deluxetable*}

\subsection{Metallicity Distributions of the Streams}\label{sec:MDF}

\begin{figure*}[ht!]
\figurenum{12}
\plotone{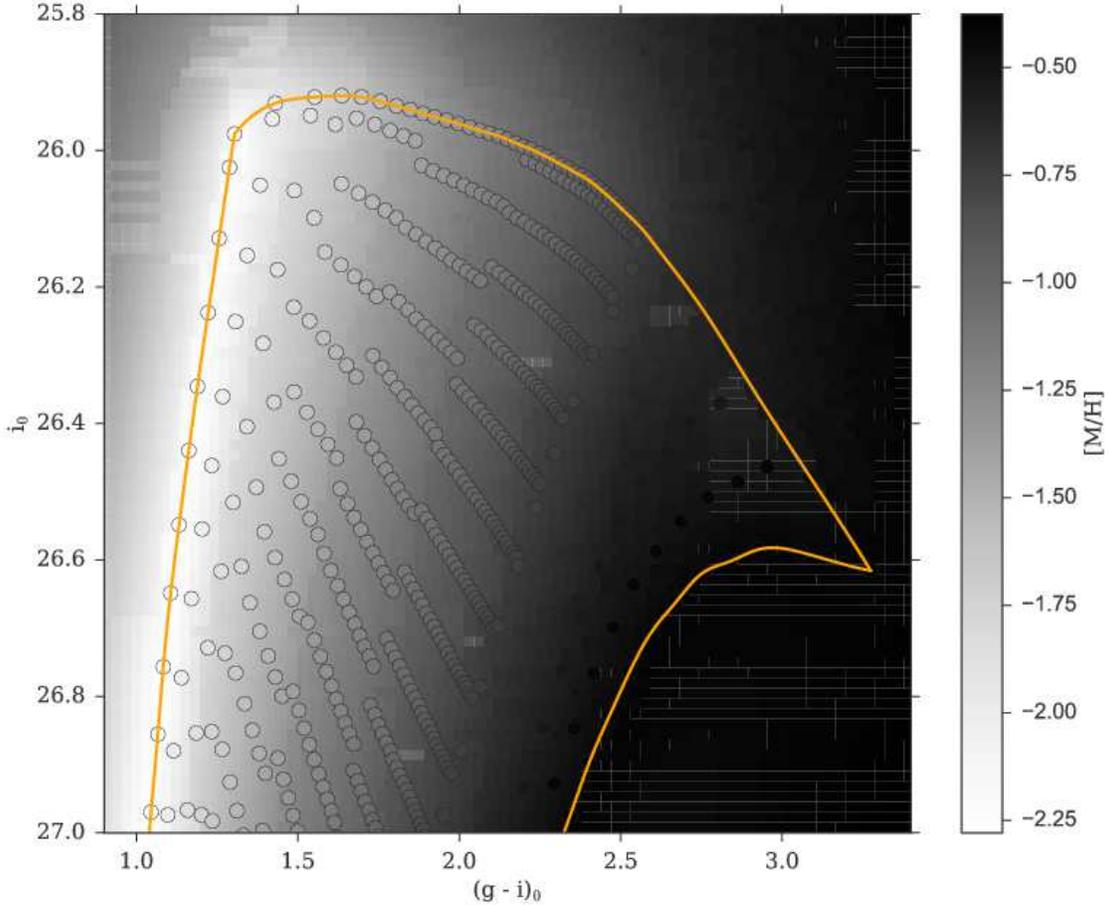}
\caption{Interpolated metallicity map on the CMD for Stream~NW. The orange solid lines at the blue and red ends, respectively, show the templates of the most metal-poor ([M/H]$ = -2.28$), metal-rich isochrone ([M/H]$ = -0.38$). The line connecting the brightest ends of these lines depends on the distance to the stream determined in Section~\ref{sec:distance}. Open circles present data points of theoretical PARSEC isochrones \citep{Bressan12} used for our metallicity interpolation scheme. The color bar shows metallicity, [M/H], converted by assuming that solar metallicity is $Z_{\sun} = 0.019$.
\label{fig:cmdMDF}}
\end{figure*}

In order to construct the metallicity distributions (MDFs) for the two stellar streams, we first assume that all of stars in the CMDs are old ($t=10$ Gyr), so that the location of the RGB stars in the CMD depends almost solely on metallicity. The photometric metallicity of each star is derived from comparison with theoretical PARSEC isochrones \citep{Bressan12} on the CMD. Figure~\ref{fig:cmdMDF} shows the interpolated metallicity map on the CMD. The interpolation is conducted based on the radial basis function (Rbf) of Python/SciPy package. Then, we select the target RGB stars brighter than $i_0 = 27.0$ mag considering the uncertainty of incompleteness. Also, since metallicity conjecture outside the criteria of the orange solid lines in Figure~\ref{fig:cmdMDF} is quite uncertain, we select only the data inside the orange ones. 

\begin{figure*}[ht!]
\figurenum{13}
\plotone{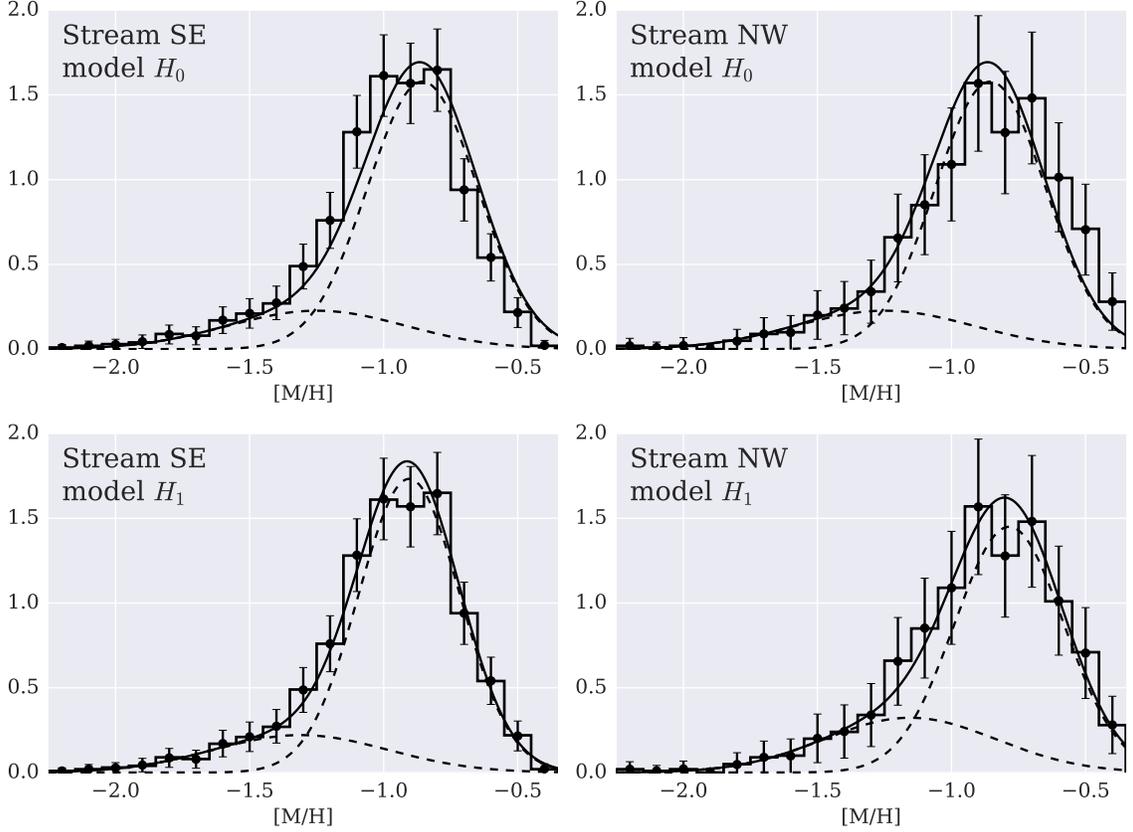}
\caption{Contaminations-subtracted, normalized MDFs of Stream~SE/NW and model $H_0$/$H_1$ (see text). Solid lines show best fit GMMs of both streams based on a maximum of posterior, while dashed lines present each component of the GMMs.\label{fig:MDF}}
\end{figure*}

The stream fields are highly contaminated by background galaxies and stars of NGC~4631's halo. Therefore, we subtract the MDF of the control field, which is a nearly flat distribution, from the MDF of the stream field. Figure~\ref{fig:MDF} shows the resultant contaminations-subtracted, normalized MDFs for the two streams. The vertical error bars denote a nominal uncertainty in each metallicity bin as derived from the Poisson errors. Both MDFs have a broad distribution ranging from the most metal-poor part of the isochrones to the most metal-rich part and there is a clear high-metallicity peak at [M/H]$ > -1$. The comprehensive shapes of both MDFs are quite similar as reproduced by Gaussian mixture model (GMM) which is a weighted sum of 2 component Gaussian densities. 

According to the $N$-body simulation of \citet{Martinez15}, both streams originated from a tidal interaction between NGC~4631 and a single dwarf satellite. If it is true, Stream~SE and NW have the same stellar population. Therefore, it is worth comparing the MDs in order to examine the validity of their model. Our approach is based on Bayesian model selection, and to do so, we prepare two models, $H_0$ and $H_1$ as follows. 

In model $H_0$, we assume that both MDFs are reproduced by the same GMM such as the following equation,
\begin{eqnarray}
x^{\rm SE}_{i} &\sim &\sum^{2}_{k=1} w_k\mathcal{N}(\mu_k,\sigma_k), \nonumber \\
x^{\rm NW}_{j} &\sim& \sum^{2}_{k=1} w_k\mathcal{N}(\mu_k,\sigma_k),
\end{eqnarray}
where $x^{\rm SE}_{i}$ and $x^{\rm NW}_{j}$ are data of Stream~SE and NW, respectively, which are re-sampled based on a probability distribution function, that is, a contaminations-subtracted, normalized MDF, and the mixture weights satisfy the constraint that $\sum^2_{k=1} w_k =1$. The likelihood then takes the form 
\begin{eqnarray}
\mathcal{L}(\bm{x}| \bm{\theta}_{H_0})&=&\mathcal{L}(\bm{x_{\rm SE}}, \bm{x_{\rm NW}}| \bm{\theta}_{H_0}) \nonumber \\
&=& \prod_{i=1}^{N_{\rm SE}} f(x^{\rm SE}_{i}|\bm{\theta}_{H_0})\prod_{j=1}^{N_{\rm NW}} f(x^{\rm NW}_{j}|\bm{\theta}_{H_0}),
\end{eqnarray}
where $\bm{\theta}_{H_0} = (w, \mu_1,\mu_2,\sigma_1,\sigma_2)$. By Bayes' theorem, we have the posterior probability distribution for the parameters, $f(\bm{\theta}_{H_0}|\bm{x}) \propto \mathcal{L}(\bm{x}| \bm{\theta}_{H_0})f(\bm{\theta}_{H_0})$, where each prior, $f(\bm{\theta}_{H_0})$, obeys visually-restricted normal distributions.

On the other hand, in model $H_1$, we assume that both MDFs are respectively reproduced by different GMM such as the following equation,
\begin{eqnarray}
x^{\rm SE}_{i} &\sim& \sum^{2}_{k=1} w^{\rm SE}_k\mathcal{N}(\mu^{\rm SE}_k,\sigma^{\rm SE}_k), \nonumber \\
x^{\rm NW}_{j} &\sim& \sum^{2}_{k=1} w^{\rm NW}_k\mathcal{N}(\mu^{\rm NW}_k,\sigma^{\rm NW}_k).
\end{eqnarray}
The likelihood function is 
\begin{eqnarray}
\mathcal{L}(\bm{x}| \bm{\theta}_{H_1}) = \prod_{i=1}^{N_{\rm SE}} f(x^{\rm SE}_{i}|\bm{\theta}^{\rm SE}_{H_1})\prod_{j=1}^{N_{\rm NW}} f(x^{\rm NW}_{j}|\bm{\theta}^{\rm NW}_{H_1}),
\end{eqnarray}
where $\bm{\theta}_{H_1} = (\bm{\theta}^{\rm SE}_{H_1}, \bm{\theta}^{\rm NW}_{H_1})$,
\begin{eqnarray}
\bm{\theta}^{\rm SE}_{H_1} &=& (w^{\rm SE},\mu^{\rm SE}_1,\mu^{\rm SE}_2,\sigma^{\rm SE}_1,\sigma^{\rm SE}_2), \nonumber \\
\bm{\theta}^{\rm NW}_{H_1} &=& (w^{\rm NW},\mu^{\rm NW}_1,\mu^{\rm NW}_2,\sigma^{\rm NW}_1,\sigma^{\rm NW}_2). \nonumber
\end{eqnarray}
By Bayes' theorem, we have the posterior probability distribution for the parameters, $f(\bm{\theta}_{H_1}|\bm{x}) \propto \mathcal{L}(\bm{x}| \bm{\theta}_{H_1})f(\bm{\theta}_{H_1})$, where each prior, $f(\bm{\theta}_{H_1})$, also obeys visually-restricted normal distributions.

In order to estimate the posterior distributions, we apply an approximate Bayesian inference, automatic differentiation variational inference (ADVI) of the PyStan package, which is known as faster algorithm than MCMC sampling \citep{Kucukelbir15}. However, the GMM inference strongly depends on initial values because the model has many local optimum and almost non-informative priors. Therefore, we explore a maximum of posterior distribution by updating initial values with higher likelihood through 1,000 iterations. Then, 20,000 samples per an iteration are made by the ADVI algorithm. Eventually, best fit GMMs of both streams based on a maximum of posterior are described by solid lines of Figure~\ref{fig:MDF}. In addition, we summarize posterior predictive distributions for the best fit GMM parameters for each model in Table~\ref{tab:posterior}. According to the estimations from model $H_1$, Stream~SE seems to be systematically more metal-poor than model $H_0$, while Stream~NW is systematically more metal-rich.

Our goal in this section is to test our hypotheses, $H_0$ and $H_1$. Here, we introduce the Watanabe-Akaike (or Widely Applicable) information criterion \citep[WAIC;][]{Watanabe10} for the model selection. It can be viewed as an improvement on the deviance information criteria (DIC) for Bayesian models \citep{Spiegelhalter02}, and one of a family of criteria that estimate the predictive power of a model, that is, how a model can anticipate new data. In addition, the better known Akaike information criterion \citep[AIC;][]{Akaike74} uses the maximum-likelihood estimate, while the WAIC averages over the posterior distribution of the parameters. Namely, the WAIC is more effective than the traditional AIC in the Bayesian approach. WAIC$_{H}$ for a model, $H$, is defined as
\begin{equation}
WAIC_{H}=-2Mean({\rm ln}(\mathcal{L}^{\ast}(\bm{x}| \bm{\theta}_{H})))+2Var({\rm ln}(\mathcal{L}^{\ast}(\bm{x}| \bm{\theta}_{H})))
\end{equation}
where $\mathcal{L}^{\ast}(\bm{x}| \bm{\theta}_{H})$ is a posterior predictive distribution for a model, $H$, that is, $H_0$ or $H_1$. As a result, WAICs of our models, $H_0$ and $H_1$, are 7271.9 and 9000.0, respectively. Therefore, we can automatically adopt the model $H_0$, and we conclude that both MDFs are reproduced by the same GMM in this study, suggesting that Stream~SE and NW have the same stellar population originated from a single dwarf satellite. Based on the model of the same stellar population, we can immediately estimate metallicity probability distribution of the progenitor from posterior predictive distribution as summarized in Table~\ref{tab:metalresult}. 

\begin{deluxetable*}{cc|rrrrrrr}
\tablecaption{Posterior Predictive Distributions of the Best Fit GMM Parameters\label{tab:posterior}}
\tablenum{4}
\tablewidth{0pt}
\tablehead{
\colhead{Model} & \colhead{$\bm{\theta}$} & \colhead{MAP} & \colhead{EAP} & \colhead{5\%} &  \colhead{25\%} & \colhead{50\%} & \colhead{75\%} & \colhead{95\%}
}
\startdata
{} & $w$ & $0.175$ & $0.189$ & $0.166$ & $0.179$ & $0.189$ & $0.199$ & $0.214$ \\
{} & $\mu_1$ & $-1.247$ & $-1.249$ & $-1.257$ & $-1.252$ & $-1.249$ & $-1.246$ & $-1.241$ \\
{$H_0$} & $\mu_2$ & $-0.861$ & $-0.855$ & $-0.869$ & $-0.861$ & $-0.855$ & $-0.850$ & $-0.841$ \\
{} & $\sigma_1$ & $0.323$ & $0.333$ & $0.308$ & $0.322$ & $0.332$ & $0.343$ & $0.358$ \\
{} & $\sigma_2$ & $0.205$ & $0.205$ & $0.197$ & $0.202$ & $0.205$ & $0.208$ & $0.213$ \\
\hline
{} & $w^{\rm SE}$ & $0.177$ & $0.181$ & $0.148$ & $0.166$ & $0.180$ & $0.194$ & $0.217$ \\
{} & $\mu^{\rm SE}_1$ & $-1.309$ & $-1.310$ & $-1.322$ & $-1.315$ & $-1.310$ & $-1.305$ & $-1.298$ \\
{} & $\mu^{\rm SE}_2$ & $-0.915$ & $-0.905$ & $-0.923$ & $-0.912$ & $-0.905$ & $-0.897$ & $-0.886$ \\
{} & $\sigma^{\rm SE}_1$ & $0.341$ & $0.326$ & $0.286$ & $0.308$ & $0.325$ & $0.342$ & $0.368$ \\
{$H_1$} & $\sigma^{\rm SE}_2$ & $0.190$ & $0.189$ & $0.179$ & $0.185$ & $0.189$ & $0.192$ & $0.198$ \\
{} & $w^{\rm NW}$ & $0.248$ & $0.261$ & $0.223$ & $0.244$ & $0.260$ & $0.276$ & $0.300$ \\
{} & $\mu^{\rm NW}_1$ & $-1.159$ & $-1.155$ & $-1.170$ & $-1.161$ & $-1.155$ & $-1.148$ & $-1.139$ \\
{} & $\mu^{\rm NW}_2$ & $-0.789$ & $-0.785$ & $-0.809$ & $-0.795$ & $-0.785$ & $-0.776$ & $-0.761$ \\
{} & $\sigma^{\rm NW}_1$ & $0.317$ & $0.321$ & $0.293$ & $0.309$ & $0.321$ & $0.333$ & $0.351$ \\
{} & $\sigma^{\rm NW}_2$ & $0.200$ & $0.203$ & $0.193$ & $0.199$ & $0.203$ & $0.208$ & $0.214$ \\
\enddata
\end{deluxetable*}

\begin{deluxetable*}{c|ccccccc}
\tablecaption{Metallicity Probability Distribution of the Progenitor\label{tab:metalresult}}
\tablenum{5}
\tablewidth{0pt}
\tablehead{
\colhead{} & \colhead{MAP} & \colhead{EAP} & \colhead{5\%} &  \colhead{25\%} & \colhead{50\%} & \colhead{75\%} & \colhead{95\%}
}
\startdata
[M/H] & $-0.85$ & $-0.92$ & $-1.46$ & $-1.07$ & $-0.89$ & $-0.73$ & $-0.51$
\enddata
\tablecomments{The 90\% (50\%) credible intervals for the metallicity is [$-1.46$, $-0.51$] ([$-1.07$, $-0.73$]). Furthermore, the negative $skewness \ (-0.80)$ indicates that the tail on the metal-poor side of the MDF is longer and broader than the metal-rich side.}
\end{deluxetable*}

\section{Discussion and Concluding Remarks}\label{sec:conclusion}

In this paper, we report the first results of the Subaru/HSC survey of the interacting galaxy system consisting of NGC~4631 and NGC~4656. We have conducted the careful analysis for our HSC data based on the HSC pipeline and our own sky-subtraction procedure. Eventually, we have detected 11 dwarf galaxies (including already-known dwarfs) in the outer region of NGC~4631 and the two tidal stellar streams around NGC~4631 in the stellar density maps divided into each resolved stellar population of MS, RSG, AGB and RGB. To discuss individual properties of these dwarf galaxies is beyond the scope of this paper, hence it will be described in the forthcoming paper. 

In this paper, we particularly focus on the fundamental properties of the two tidal streams. Based on the TRGB method and the Bayesian statistics, we find that Stream~SE (at the heliocentric distance of 7.10~Mpc in EAP with the 90\% credible intervals of [6.22, 7.29]~Mpc) is located in front of NGC~4631 along the line-of-sight, while Stream~NW (at 7.91~Mpc in EAP with the 90\% credible intervals of [6.44, 7.97]~Mpc) is nestled behind NGC~4631. On the other hand, we calculate metallicity distribution of each stream by comparing each star with theoretical isochrones on the color-magnitude diagram (CMD), and we find the possibility that both streams have the same stellar population based on the Bayesian model selection method, suggesting that they originated from a tidal interaction between NGC~4631 and a single dwarf satellite. 

The expected progenitor has a positively skewed MDF with $\rm [M/H]_{\rm EAP}=-0.92$ with the 90\% credible intervals of [$-1.46, -0.51$]. Provided that $\rm [M/H]=[Fe/H]+log(0.694\times[\alpha/Fe] + 0.306)$ \citep{CS13} and $\rm [\alpha/Fe]=0.3$ ($\alpha$-enhancement of old stellar populations of the Milky Way), we estimate the mass of the progenitor as $3.7 \times 10^8 M_\sun$ with the 90\% credible intervals of [$5.8 \times 10^6, 8.6 \times 10^9$]~$M_\sun$ based on the mass-metallicity relation for Local group dwarf galaxies \citep{Kirby13}. This is in good agreement with a total initial stellar mass ($5.4 \times 10^8 M_\sun$) of the progenitor presumed in the $N$-body simulation of \citet{Martinez15}. However, red/metal-rich ends of these MDFs suffer from incompleteness even if we carefully construct them by considering completeness function derived in Section~\ref{sec:completeness}. Therefore we cannot exclude that the true MDFs of the streams may be more metal-rich if such red stars are present in these fields. This also suggests that the mass of the progenitor galaxy estimated in this study may be a lower limit. 

To compare with other resolved stellar streams in galaxies of the local universe, we calculate the surface brightness of Stream~SE and NW. First, we extract the secure RGB stars for both the two streams and each control field within orange lines shown in Figure~\ref{fig:cmdMDF}. In this procedure, we adopt the lower limit of $i$-band magnitude to 1.5 mag fainter than each TRGB magnitude estimated in Section~\ref{sec:distance}. Second, we convert the summed-up flux counts of selected stars to the surface brightness in mag arcsec$^{-2}$, and subtract the surface brightness of the control field, for which remaining foreground and background contaminations are removed based on the statistical method as discussed in previous sections. To convert the surface brightness in HSC filter systems to the one in standard Johnson $V$-band, we adopt the following formula (Komiyama et al. 2017, in preparation):
\begin{equation}
g-V=0.371(g-i)+0.068.
\end{equation}
Finally, we measure the mean surface brightnesses within the red-dashed circles of Figure~\ref{fig:plot_stream} of $\langle \mu_{V} \rangle = 31.0\pm0.02$ mag~arcsec$^{-2}$ for Stream~SE and $\langle \mu_{V} \rangle = 32.2\pm0.02$ mag~arcsec$^{-2}$ for Stream~NW. The errors are estimated from the Poisson statistics from the finite number of observed stars and subtracted contaminations. Although Stream~SE and NW are the brightest stellar streams in NGC~4631's system, they may be relatively faint in comparison with other resolved stellar streams detected in galaxies of the local universe according to observational evidence of the relation between metallicity and surface brightness of stellar substructures of M31 and NGC~55 \citep{Gilbert09,Tanaka11}. 

We examine the spatial structures of surface brightness of the two streams and find that the brightest regions of Stream~SE and NW are located at $(\alpha,\delta)\sim(190.75\arcdeg,32.38\degr)$ and $(\alpha,\delta)\sim(190.24\arcdeg,32.79\degr)$, and their surface brightnesses within 0.01 deg$^2$ bin corresponding to 1.4~kpc$^2$ and 1.2~kpc$^2$ are $\mu_{\rm V} = 30.1$ mag~arcsec$^{-2}$ and $\mu_{\rm V} = 29.5$ mag~arcsec$^{-2}$, respectively. Although the errors estimated from Poisson noise are less than 0.01 mag~arcsec$^{-2}$, the values of these surface brightness are probably lower limits due to the uncertainties of incompleteness corrections of detection and blending. Their positions are consistent with density peaks of the KDE maps of Figure~\ref{fig:stream_SE} and \ref{fig:stream_NW}. Therefore, Stream~NW has a brighter and more concentrated core than Stream~SE, implying that the main body of the progenitor is probably associated with Stream~NW rather than Stream~SE. In that case, it is reasonably thought that Stream~SE formed through a tidal interaction between a dwarf satellite embedded in Stream~NW and NGC~4631 several Gyrs ago. This interpretation is consistent with the prediction by the $N$-body simulation of \citep{Martinez15}. On the other hand, a continuous, bridge-like structure between Stream~SE and NW that the model predicts are not detected in this study. If such a faint structure exists, its surface brightness is less than background, that is, $\mu_{\rm V} = 32.8 \pm 0.3$ mag~arcsec$^{-2}$ in the control field for Stream~SE and $\mu_{\rm V} = 33.1 \pm 0.3$ mag~arcsec$^{-2}$ in the control field for Stream~NW. These errors indicate standard deviations of the surface brightness within the total bins in each control field. Furthermore, the projected length of that structure is $\sim 100$~kpc, as measured between the cores of the two stream components, which is consistent with that simulation. Meanwhile, that structure might have the large extent more than several hundred kpc along the line-of-sight.

Currently, there is observational evidence that stellar halos may become less common at lower stellar masses than Milky Way mass spiral galaxies \citep[e.g.,][]{Tanaka11,Streich16} although there is a variation in the masses of stellar halos of spiral galaxies with stellar masses similar to that of the Milky Way \citep{Merritt16,Harmsen16}. NGC~4631 is interpreted as a large Magellanic-type galaxy for many years, and it is classified as a relatively late-type spiral galaxy in the Local Volume \citep{deVaucouleurs63}. Therefore, we can infer that NGC~4631 is less massive galaxy than the Milky Way. Notwithstanding, NGC~4631 has a relatively active accretion history that continues to influence growth of its stellar halo characterized by the two large tidal stellar streams. It is expected that the further analysis of its stellar halo constrains formation mechanism of low mass stellar halos.

\acknowledgments

We thank the staff at the Subaru telescope for their excellent support during our observing runs. We wish to recognize and acknowledge the very significant cultural role and reverence that the sacred summit of Mauna a Wakea, which is the dwelling place of the goddess Poli'ahu, has always had within the endemic Hawaiian community. We are the most fortunate to have the opportunity to conduct observations from this sacred mountain. We are grateful to the referee for helpful comments and suggestions. This work was supported by JSPS KAKENHI Grant Number JP25800098 for MT and JP15K05037 for YK. This work was supported in part by MEXT Grant-in-Aid for Scientific Research on Innovative Areas (No.JP15H05889, JP16H01086 for MC). 

\appendix

\section{technique of sky subtraction}

\begin{figure}[ht!]
\figurenum{14}
\plotone{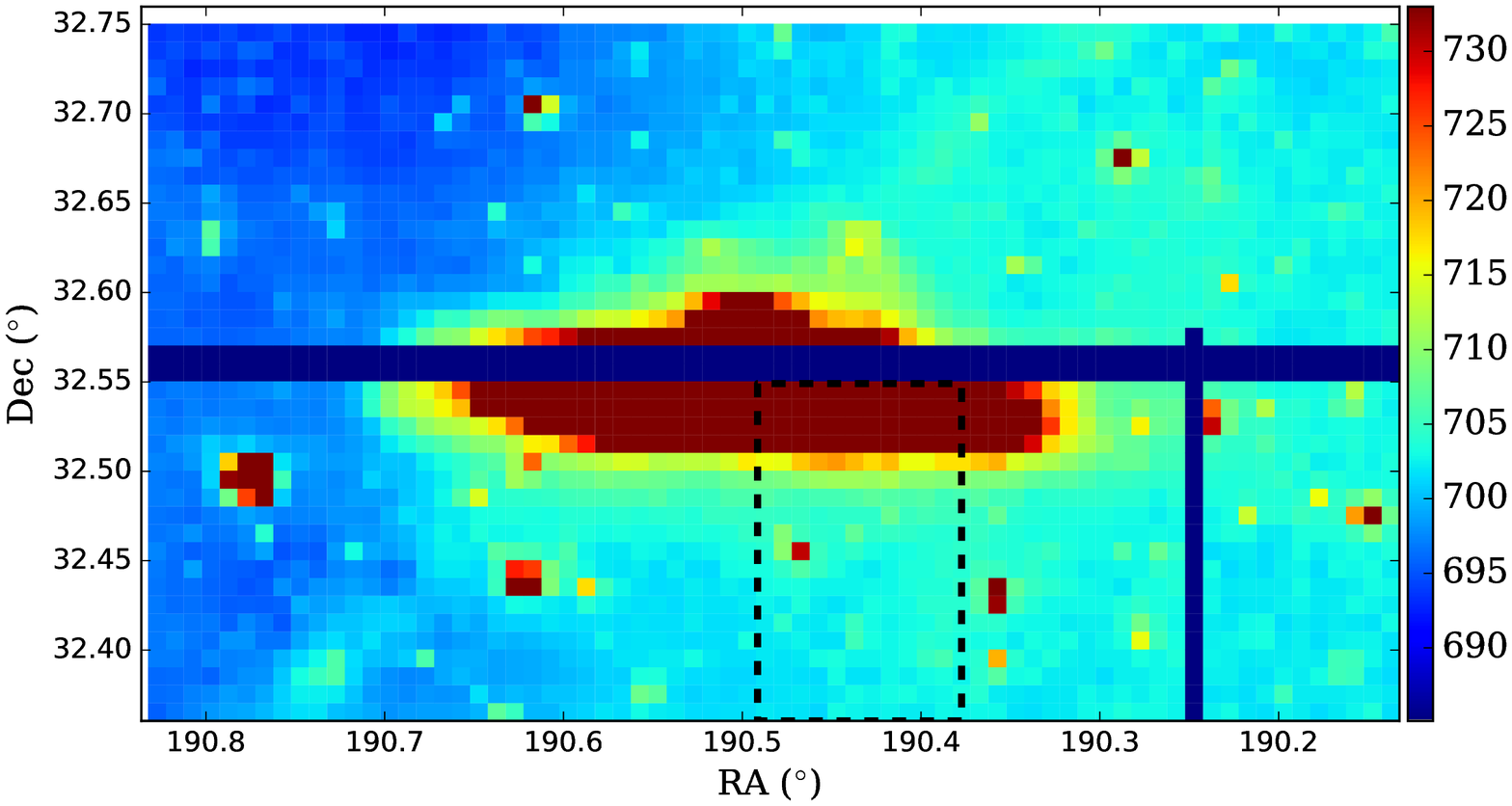}
\plotone{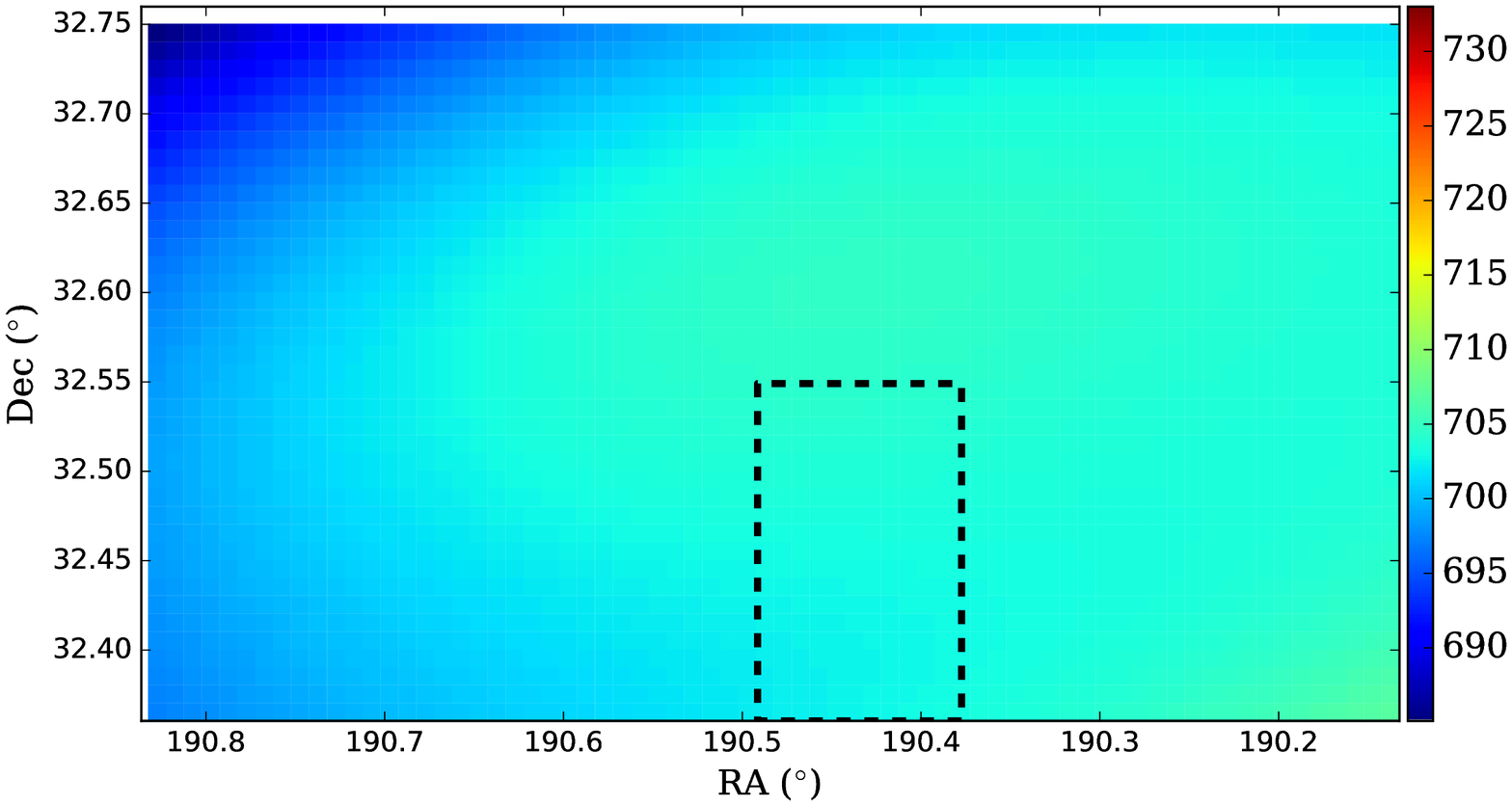}
\caption{{\it Top}: a color map of unmasked flat-fielded data consisting 10 ccd chips surrounding NGC~4631's main body. The right side bar shows count values in each bin with $0.01 \times 0.01$ square degrees. {\it Bottom}: a color map of sky-gradient model fitted by the fourth order Chebyshev polynomial. The black dashed rectangles of both panels are corresponding to the CCD chip of Figure~\ref{fig:skysubtraction}.\label{fig:skymap}}
\end{figure}

\begin{figure}[ht!]
\figurenum{15}
\plotone{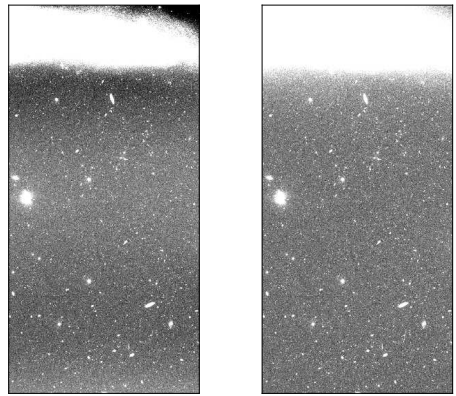}
\caption{Comparison of the sky-subtraction result of the HSC pipeline (left) with that of our method (right), focusing on one CCD chip with number 065. The reduced image based on the HSC pipeline is apparently oversubtracted on the periphery of the large object.\label{fig:skysubtraction}}
\end{figure}

The raw data were basically reduced with HSC pipeline 3.8.5. We used SDSS Data Release 7 as an astrometric catalog in the pipeline. However, the sky-subtraction process of the pipeline didn't work well in regard to CCD chips suffered from an apparently large object such as nearby galaxies (see also the left panel of Figure~\ref{fig:skysubtraction}). Therefore, we applied our own sky-subtraction method to such contaminated CCD chips. The top panel of Figure~\ref{fig:skymap} shows a color map of unmasked flat-fielded data consisting 10 ccd chips surrounding NGC~4631's main body. First, we masked large objects, that is, NGC~4631 and NGC~4656, in flat-fielded data considering their coordinates, major-axis and minor-axis sizes based on NED as well as saturated/bad pixels based on mask information made from the HSC pipeline and ones with higher signal-to-noise ratio than $2\sigma$. Second, we conducted binning with 128 pixels for the data due to the limitation of our computational power. Third, assuming that the masked flat-fielded data is corresponding to natural sky gradients, we reconstructed sky-gradient model (see the bottom panel of Figure~\ref{fig:skymap}) using the fourth order Chebyshev polynomial. 

\begin{figure}[ht!]
\figurenum{16}
\plotone{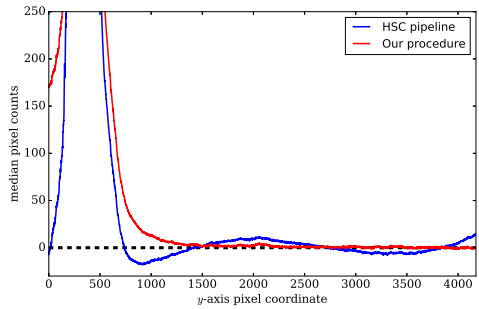}
\caption{Spacial variations of median count values of the same sky-subtracted CCD chips as Figure~\ref{fig:skysubtraction}. The red line shows the data reduced by our procedure, while the blue one presents that reduced by the HSC pipeline.\label{fig:skysubtraction_median}}
\end{figure}

The right hand panel of Figure~\ref{fig:skysubtraction} shows the sky-subtracted image based on our sky-gradient model, suggesting that our sky-subtraction process works better than that of the HSC pipeline. In order to statistically compare the sky-subtracted data by the HSC pipeline with the one by our procedure, we investigate how pixel counts (median values stacked along the $x$-axis) along the $y$-axis of the 2k $\times$ 4k CCD chips of Figure~\ref{fig:skysubtraction} change. Figure~\ref{fig:skysubtraction_median} shows that the data reduced by our procedure is reaching to no background gradient around zero with increasing distance from NGC~4631's main body, while the data reduced by the HSC pipeline is strongly swelling with highly-deviated background. On the other hand, sky of unsuffered CCD chips was subtracted using the sky models estimated in each CCD chip based on our sky-subtraction procedure.

\begin{thebibliography}{57}
\bibitem[Akaike (1974)]{Akaike74}
Akaike, H.\ 1974, ITAC, 19, 716
\bibitem[Axelrod et al.(2010)]{Axelrod10}
Axelrod, T., Kantor, J., Lupton, R.~H., \& Pierfederici, F.\ 2010, Proc. SPIE, 7740, 774015
\bibitem[Bellazzini (2008)]{Bellazzini08}
Bellazzini, M.\ 2008, MmSAI, 79, 440
\bibitem[Bressan et al.(2012)]{Bressan12}
Bressan, A., Marigo, P., Girardi, L., et al.\ 2012, \mnras, 427, 127
\bibitem[Bullock \& Johnston(2005)]{BJ05}
Bullock, J.~S., \& Johnston, K.~V.\ 2005, \apj, 635, 931
\bibitem[Cassisi \& Salaris(2013)]{CS13}
Cassisi S., \& Salaris M.\ 2013, Old Stellar Populations: How to Study the Fossil Record of Galaxy Formation
\bibitem[Conn et al.(2011)]{Conn11}
Conn, A.~R., Lewis, G.~F., Ibata, R.~A., et al.\ 2011, \apj, 740, 69
\bibitem[Conn et al.(2012)]{Conn12}
Conn, A.~R., Ibata, R.~A., Lewis, G.~F., et al.\ 2012, \apj, 758, 11
\bibitem[Cooper et al.(2010)]{Cooper10}
Cooper, A.~P., Cole, S., Frenk, C.~S., et al.\ 2010, \mnras, 406, 744
\bibitem[Crnojevi\'c et al.(2016)]{Crnojevic16}
Crnojevi\'c, D., et al.\ 2016, \apj, 823, 19
\bibitem[de Vaucouleurs \& de Vaucouleurs(1963)]{deVaucouleurs63}
de~Vaucouleurs, G., \& de~Vaucouleurs, A.\ 1963, \apj, 137, 363
\bibitem[Fleming et al.(1995)]{Fleming95}
Fleming, D.~E.~B., Harris, W.~E., Pritchet, C.~J., Hanes, D.~A.\ 1995, \aj, 109, 1044
\bibitem[Gilbert et al.(2009)]{Gilbert09}
Gilbert, K.~M., Font, A.~S., Johnston, K.~V., \& Guhathakurta, P.\ 2009, \apj, 701, 776
\bibitem[Greggio et al.(2014)]{Greggio14}
Greggio, L., Rejkuba, M., Gonzalez, O.~A., Arnaboldi, M., Iodice, E., Irwin, M., Neeser, M.~J., and Emerson, J.\ 2014, \aap, 562, 73
\bibitem[Harmsen et al.(2016)]{Harmsen16}
Harmsen, B., Monachesi, A., Bell, E.~F., de~Jong, R.~S., Bailin, J., Radburn-Smith, D.~J., \& Holwerda, B.~W.\ 2017, \mnras, 466, 1491
\bibitem[Ibata et al.(1994)]{Ibata94}
Ibata, R.~A., Gilmore, G., \& Irwin, M.~J.\ 1994, \nat, 370, 194
\bibitem[Ibata et al.(2001)]{Ibata01}
Ibata, R.~A., et al.\ 2001, \nat, 412, 49
\bibitem[Ibata et al.(2007)]{Ibata07}
Ibata, R.~A., Martin, N.~F., Irwin, M., et al.\ 2007, \apj, 671, 1591
\bibitem[Ibata et al.(2014)]{Ibata14}
Ibata, R.~A., Lewis, G.~F., McConnachie, A.~W., et al.\ 2014, \apj, 780, 128
\bibitem[Javanmardi et al.(2016)]{Javanmardi16}
Javanmardi, B., Mart\'inez-Delgado, D., Kroupa, P., et al.\ 2016, \aap, 588, 89
\bibitem[Johnston et al.(2008)]{Johnston08}
Johnston, K.~V., et al.\ 2008, \apj, 689, 936
\bibitem[Karachentsev et al.(2014)]{Karachentsev14}
Karachentsev, I.~D., Bautzmann, D., Neyer, F., et al.\ 2014, astro-ph/1401.2719
\bibitem[Kirby et al.(2013)]{Kirby13}
Kirby, E.~N., Cohen, J.~G., Guhathakurta, P., et al.\ 2013, \apj, 779, 102
\bibitem[Kucukelbir et al.(2015)]{Kucukelbir15}
Kucukelbir, A., Ranganath, R., Gelman, A., Blei, M.~D.\ 2015, stat/1506.03431
\bibitem[Lee et al.(1993)]{Lee93}
Lee, M.~G., Freedman, W.~L., \& Madore, B.~F.\ 1993, \apj, 417, 553
\bibitem[Madore \& Freedman(1995)]{Madore95}
Madore, B.~F., \& Freedman, W.~L.\ 1995, \aj, 109, 1645
\bibitem[Majewski et al.(2003)]{Majewski03}
Majewski, S.~R., Skrutskie, M.,Weinberg, M., \& Ostheimer, J.\ 2003, \apj, 599, 1082
\bibitem[Makarov et al.(2006)]{Makarov06}
Makarov, D., Makarova, L., Rizzi, L., et al.\ 2006, \aj, 132, 2729
\bibitem[Mart\'inez-Delgado et al.(2010)]{Martinez10}
Mart\'inez-Delgado, D., Gabany, R.~J., Crawford, K., et al.\ 2010, \aj, 140, 962
\bibitem[Mart\'inez-Delgado et al.(2015)]{Martinez15}
Mart\'inez-Delgado, D., D'Onghia, E., Chonis, T.~S., et al.\ 2015, \aj, 150, 116
\bibitem[M\`endez et al.(2002)]{Mendez02}
M\`endez, B., Davis, M., Moustakas, J., et al.\ 2002, \aj, 124, 213
\bibitem[Merritt et al.(2016)]{Merritt16}
Merritt, A., van~Dokkum, P., Abraham, R., \& Zhang, J.\ 2016, \apj, 830, 62
\bibitem[Miyazaki et al.(2012)]{Miyazaki12}
Miyazaki, S., Komiyama, Y., Nakata, H., et al.\ 2012, Proc. SPIE, 8446, 84460Z
\bibitem[Monachesi et al.(2013)]{Monachesi13}
Monachesi A. et al.\ 2013, \apj, 766, 106
\bibitem[Monachesi et al.(2016)]{Monachesi16}
Monachesi A. et al.\ 2016, \mnras, 457, 1419
\bibitem[Mouhcine et al.(2010)]{Mouhcine10}
Mouhcine, M., Ibata, R., \& Rejkuba, M.\ 2010, \apj, 714, L12 
\bibitem[Okamoto et al.(2015)]{Okamoto15}
Okamoto, S., Arimoto, N., Ferguson, A.~M.~N., et al.\ 2015, \apjl, 809, L1
\bibitem[Radburn-Smith et al.(2011)]{Radburn-Smith11}
Radburn-Smith, D.~J., et al\ 2011, \apjs, 195, 18
\bibitem[Rand (1994)]{Rand94}
Rand, R.~J.,\ 1994, \aap, 285, 833
\bibitem[Robin et al.(2003)]{Robin03}
Robin, A., et al.\ 2003, \aap, 409, 523
\bibitem[Sakai et al.(1996)]{Sakai96}
Sakai, S., Madore, B.~F., \& Freedman, W.~L.\ 1996, \apj, 461, 713
\bibitem[Salaris \& Cassisi(2005)]{Salaris05}
Salaris, M., \& Cassisi, S.\ 2005, Evolution of Stars and Stellar Populations (New York: Wiley)
\bibitem[Schechtman-Rook \& Hess (2012)]{SH12}
Schechtman-Rook, A., \& Hess, K.~M.\ 2012, \apj, 750, 171
\bibitem[Schlafly \& Finkbeiner(2011)]{Schlafly11}
Schlafly, E.~F., \& Finkbeiner, D.~P.\ 2011, \apj, 737, 103 
\bibitem[Schlegel et al.(1998)]{Schlegel98}
Schlegel, D.~J., Finkbeiner, D.~P., \& Davis, M.\ 1998, \apj, 500, 525
\bibitem[Seth et al.(2005a)]{Seth05a}
Seth, A.~C., Dalcanton, J.~J., \& de~Jong, R.~S.\ 2005a, \aj, 129, 1331
\bibitem[Seth et al.(2005b)]{Seth05b}
Seth, A.~C., Dalcanton, J.~J., \& de~Jong, R.~S.\ 2005b, \aj, 130, 1574
\bibitem[Silverman (1986)]{Silverman86}
Silverman, B.~W.\ 1986, ``Density Estimation for Statistics and Data Analysis'', Vol. 26, Monographs on Statistics and Applied Probability, Chapman and Hall, London.
\bibitem[Spiegelhalter et al.(2002)]{Spiegelhalter02}
Spiegelhalter, J.~D., Best, G.~N., Carlin, P.~B., Linde, A.\ 2002, Journal of Royal Statistical Society, Series B, 64, 583
\bibitem[Streich et al.(2016)]{Streich16}
Streich, D., de~Jong, R.~S., Bailin, J., Bell, E.~F., Holwerda, B.~W., Minchev, I., Monachesi, A., Radburn-Smith, D.~J.\ 2016, \aap, 585, A97
\bibitem[Tanaka et al.(2010)]{Tanaka10}
Tanaka, M., Chiba, M., Komiyama, Y., et al\ 2010, \apj, 708, 1168
\bibitem[Tanaka et al.(2011)]{Tanaka11}
Tanaka, M., Chiba, M., Komiyama, Y., et al.\ 2011, \apj, 738, 150
\bibitem[Tikhonov et al.(2006)]{Tikhonov06}
Tikhonov, N.~A., Galazutdinova, O.~A., Drozdovsky, I.~O.\ 2006, astro-ph/0603457
\bibitem[Tollerud et al. (2016)]{Tollerud16}
Tollerud, E.~J., Geha, M.~C., Grcevich, J., et al.\ 2016, \apj, 827, 89
\bibitem[VandenBerg et al.(2006)]{VandenBerg06}
VandenBerg, D.~A., Bergbusch, P.~A., \& Dowler, P.~D.\ 2006, \apjs, 162, 375
\bibitem[van~Dokkum et al.(2014)]{vanDokkum14}
van~Dokkum, P.~G., Abraham, R., Merritt, A.\ 2014, \apjl, 782, 24L
\bibitem[Watanabe (2010)]{Watanabe10}
Watanabe, S.\ 2010, Journal of Machine Learning Research, 11, 3571
\end{thebibliography}
\end{document}